\documentclass{article}
\PassOptionsToPackage{numbers,sort&compress}{natbib}
\usepackage[final]{neurips_2024}
\usepackage[utf8]{inputenc} 
\usepackage[T1]{fontenc}    
\usepackage{hyperref}       
\usepackage{url}            
\usepackage{booktabs}       
\usepackage{amsfonts}       
\usepackage{amsmath}
\usepackage{amssymb} 
\usepackage{stmaryrd}
\usepackage[scaled=0.75]{beramono}
\usepackage{needspace}

\usepackage{nicefrac}       
\usepackage{microtype}      
\usepackage{xcolor}         
\usepackage{algorithm}
\usepackage[noend]{algpseudocode}
\usepackage{mathtools}
\usepackage{multirow}
\usepackage{tcolorbox}

\usepackage{color, soul}
\definecolor{grey}{rgb}{0.9,0.9,0.9}
\usepackage{balance}
\usepackage{textcomp}

\usepackage{alltt}
\usepackage{latexsym}
\usepackage{xspace}
\usepackage{url}
\usepackage{paralist}
\usepackage{pdfpages}
\usepackage{listings} 
\usepackage{alltt}
\usepackage{hyperref}
\usepackage{booktabs}  
\usepackage{subcaption} 
\usepackage{url}
\usepackage{hhline}
\usepackage{multicol}
\usepackage{makecell}
\usepackage{todonotes}
\usepackage[font={small}]{caption}
\usepackage{color}
\usepackage[export]{adjustbox}
\usepackage{natbib}
\usepackage[inline,shortlabels]{enumitem}
\usepackage{longtable}
\usepackage{titlesec}

\title{\tool: Context-Free LLM Approximation for Guiding Program Synthesis}
\author{%
  Shraddha Barke\\
  UC San Diego\\
  San Diego, USA\\
  \texttt{sbarke@ucsd.edu} \\
  \And
  Emmanuel Anaya Gonzalez\\
  UC San Diego\\
  San Diego, USA\\
  \texttt{fanayagonzalez@ucsd.edu} \\
  \And
  Saketh Ram Kasibatla\\
  UC San Diego\\
  San Diego, USA\\
  \texttt{skasibatla@ucsd.edu} \\
  \And
  Taylor Berg-Kirkpatrick\\
  UC San Diego\\
  San Diego, USA\\
  \texttt{tbergkirkpatrick@ucsd.edu} \\
  \And
  Nadia Polikarpova\\
  UC San Diego\\
  San Diego, USA\\
  \texttt{npolikarpova@ucsd.edu} \\
}

\definecolor{skcolor}{rgb}{0.1,0.7,0.8}
\definecolor{npcolor}{RGB}{255, 0, 255}
\definecolor{sbcolor}{RGB}{0,100,50}
\definecolor{eacolor}{rgb}{0.7,0.3,0.7}

\titlespacing*{\paragraph}{0pt}{0pt}{1em}

\def\suchthatt{\,\middle|\,}

\newcommand{\set}[2]{\left\lbrace\,#1 \suchthatt #2\,\right\rbrace}

\newcommand{\scode}[1]{{\texttt{\small #1}}}

\newcommand{\stringbench}{\tname{String}}
\newcommand{\tensorbench}{\tname{Tensor}}

\newcommand{\tool}{\textsc{HySynth}\xspace}
\newcommand{\tname}[1]{\textsc{#1}\xspace}

\newcommand{\sygus}{\tname{SyGuS}}
\newcommand{\tfcoder}{\tname{TFCoder}}
\newcommand{\arc}{\tname{Arc}}
\newcommand{\ocarc}{\tname{Object-Arc}}
\newcommand{\arga}{\tname{Arga}}

\newcommand{\probe}{\tname{Probe}}

\newcommand{\tensorflow}{TensorFlow\xspace}
\newcommand{\deepseek}{\tname{DeepSeek}}
\newcommand{\gptfouro}{\tname{Gpt4o}}
\newcommand{\gptthree}{\tname{Gpt3.5}}

\newcommand{\etc}{\emph{etc}\xspace}
\newcommand{\ie}{\emph{i.e.\@}\xspace}

\newcommand{\eg}{\emph{e.g.\@}\xspace}

\newcommand{\wrt}{\emph{wrt.\@}\xspace}

\newcommand{\astar}{A*\xspace}

\newcommand{\nsol}{\ensuremath{N}} 

\newcommand{\examples}{\mathcal{E}}
\newcommand{\grammar}{\mathcal{G}}

\newcommand{\sem}[1]{
	\llbracket
	#1 
	\rrbracket
}

\newcommand{\exec}{\ensuremath{\textsc{E}}}
\newcommand{\bank}{\ensuremath{\textsc{B}}}
\newcommand{\eval}{\ensuremath{\textsc{Eval}}}
\newcommand{\level}{\ensuremath{\textsc{Lvl}}}

\newcommand{\llmSols}{\ensuremath{\mathcal{S}_{\text{llm}}}}
\newcommand{\costfn}{\icost}

\renewcommand{\gets}{\leftarrow}

\newcommand{\transform}{T}

\newcommand{\round}[1]{\ensuremath{\lceil#1\rceil}}
\newcommand{\nontermset}{\mathcal{N}}

\newcommand{\mset}{\mathcal{M}}
\newcommand{\oset}{\mathcal{O}}
\newcommand{\ruleset}{\mathcal{R}}
\newcommand{\start}{\mathcal{S}}
\newcommand{\termset}{\Sigma}
\newcommand{\rl}{\textsc{R}}
\newcommand{\nterm}{\textsc{N}}
\newcommand{\icost}{\mathrm{cost}}
\newcommand{\rcost}{\mathrm{cost}_{\mathbb{R}}}

\newcommand{\rwei}{w_{\mathbb{R}}}
\newcommand{\wei}{w}

\newcommand{\pcfg}{p}
\newcommand{\llim}{\textsc{Lim}}
\newcommand{\prog}{P}

\algrenewcommand\algorithmicrequire{\textbf{Input:}}
\algrenewcommand\algorithmicensure{\textbf{Output:}}

\xspace
\xspace

\newcommand{\many}[1]{\overrightarrow{#1}}

\newcommand{\nonterm}[1]{\ensuremath{\mathit{#1}}}

\newcommand{\step}{\Rightarrow}
\newcommand{\manystep}{\Rightarrow^*}
\newcommand{\lang}{\mathcal{L}}
\newcommand{\vals}{\mathrm{Val}}
\newcommand{\trace}{\mathrm{tr}}

\definecolor{commentgreen}{rgb}{0.25,0.5,0.35}

\lstdefinelanguage{arc}{
	morekeywords={if,then,self},
	literate=%
    {->}{$\rightarrow$}2
    {=.=}{$\doteq$}1
    {==}{$=$}1
    {!=}{$\neq$}1
    {&&}{$\land$}1
    {||}{$\lor$}1
    {<}{$<$}1
    {>}{$>$}1
    {<=}{$\le$}1
    {>=}{$\ge$}1
	,
  morecomment=[l]{//},
	numbers=none,
	basicstyle=\ttfamily,
	commentstyle=\itshape\color{commentgreen},
	keywordstyle=\bfseries,
	ndkeywordstyle=\bfseries
}

\newcommand{\T}[1]{\mbox{\lstinline[language=arc,basicstyle=\ttfamily,columns=fixed]^#1^}}

\lstnewenvironment{longlisting}
    {\longtable{l}}
    {\endlongtable}

\begin{document}
\maketitle
\begin{abstract}
Many structured prediction and reasoning tasks can be framed as program synthesis problems, 
where the goal is to generate a program in a \emph{domain-specific language} (DSL) 
that transforms input data into the desired output.
Unfortunately, purely neural approaches, such as large language models (LLMs), often fail to produce fully correct programs in unfamiliar DSLs,
while purely symbolic methods based on combinatorial search scale poorly to complex problems.
Motivated by these limitations, we introduce a hybrid approach, 
where LLM completions for a given task are used to learn a task-specific, context-free surrogate model,
which is then used to guide program synthesis. 
We evaluate this hybrid approach on three domains,
and show that it outperforms both unguided search and direct sampling from LLMs,
as well as existing program synthesizers.
\end{abstract}

\section{Introduction}

\begin{figure}[t]
\centering
\begin{subfigure}[t]{.25\textwidth}
\includegraphics[width=\textwidth]{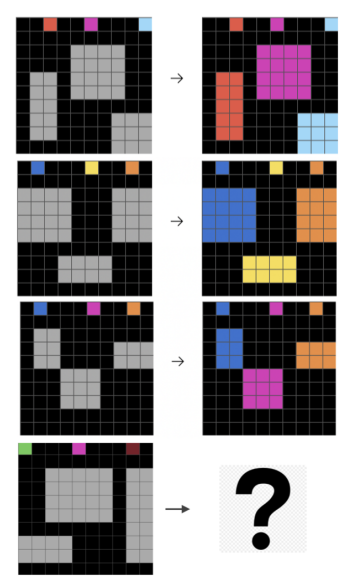}
\caption{\arc}
\label{fig:pbe-a}\end{subfigure}%
\begin{subfigure}[t]{.35\textwidth}
\includegraphics[width=\textwidth]{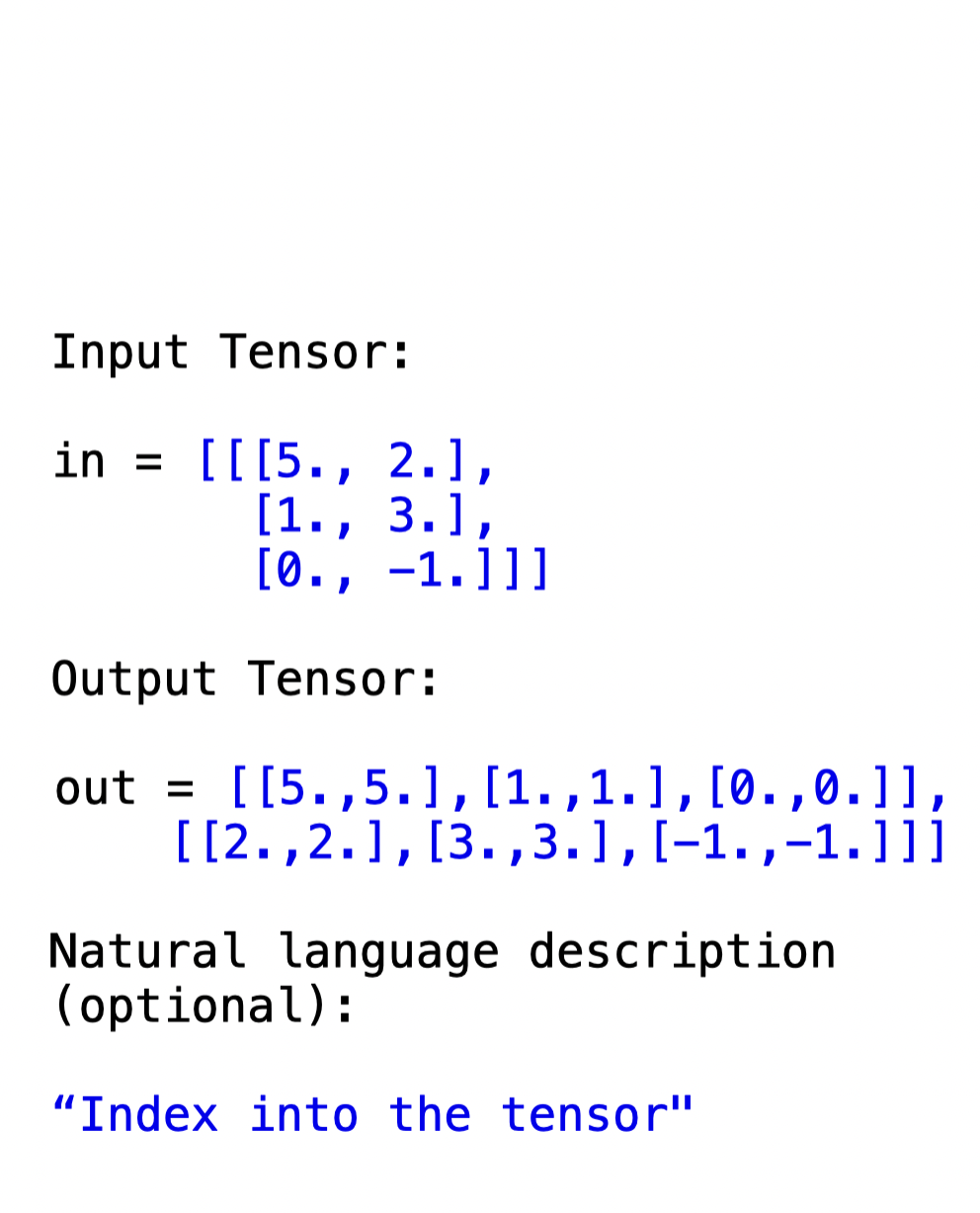}
\caption{\tensorbench}
\label{fig:pbe-b}
\end{subfigure}%
\begin{subfigure}[t]{.35\textwidth}
\includegraphics[width=\textwidth]{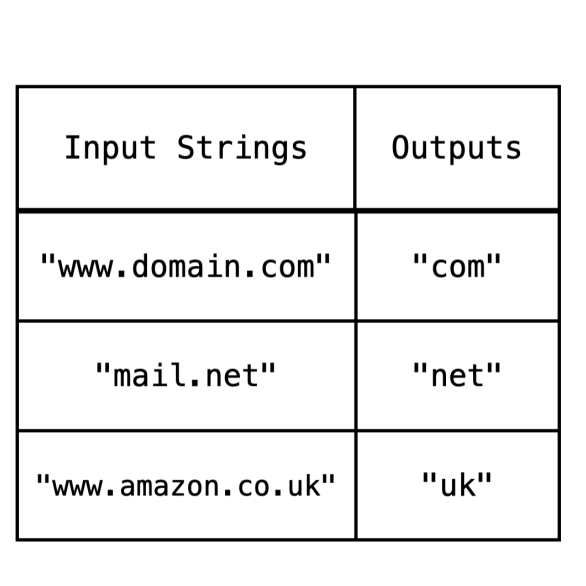}
\caption{\stringbench}
\label{fig:pbe-c}
\end{subfigure}
\caption{
	Example problems from the three PBE domains we evaluate \tool on:
	grid-based puzzles (\arc), tensor manipulation (\tensorbench), and string manipulation (\stringbench).
}
\label{fig:pbe}
\end{figure}

Large language models (LLMs) demonstrate impressive capabilities in various domains,
but they continue to struggle with tasks that require precision---e.g.~structured prediction, reasoning, counting, or data transformation---when direct task examples are not prevalent in their training data \cite{tan2023large,xu2023llms,berglund2023reversal,bai2023constituency,josifoski2023exploiting,ugare2024improving,mccoy2023embers}.
As one example, consider the \emph{Abstraction and Reasoning Corpus} (\arc)~\cite{chollet2019measure},
which was designed as a benchmark for human-like structured reasoning.
\arc tasks are grid-based puzzles, such as one depicted in \autoref{fig:pbe-a}.
This puzzle consists of three training examples, which are pairs of input and output grids;
the goal is to infer the transformation that maps the input to the output,
and then apply this transformation to the test grid.
The \arc benchmark's emphasis on generalization and few-shot learning has rendered it challenging to solve with purely machine learning techniques:
state-of-the-art generative models like GPT-4 hardly solve more than 10\% of the tasks in the dataset when asked to predict the test output, even with the help of advanced prompting techniques \cite{lee2024reasoning}.


In fact, the leading entries in the \arc Kaggle competition \cite{arc_kaggle} tackle this task using \emph{Programming-by-Example} (PBE):
instead of predicting the output directly, they search for a program
that captures the transformation occurring in the input-output examples.
%
%
For example, the transformation in \autoref{fig:pbe-a} might be represented as the following program:
\begin{multline}\label{eq:example-program}
\T{if color\_of(self) == GREY && is\_neighbor(self, other) && size\_of(other) == MIN}\\
\T{then update\_color(color\_of(other))}
\end{multline}
This particular program is written in a \emph{domain-specific language} (DSL) inspired by the \arga tool~\cite{xu2023graphs}.
It consists of a single \emph{rule} of the form \T{if} \emph{filter} \T{then} \emph{transform}, 
which is applied to each object in the grid simultaneously;
if the filter holds for the focus object \T{self} and another object \T{other},
then \T{self} undergoes the transform.
%
%
In this case, the rule says that any grey object that has a neighbor of the grid's minimum size (here, a single pixel)
should be colored with the color of that neighbor.

Beyond grid puzzles, PBE is a general paradigm for structured reasoning and data transformation tasks:
for example, it can help spreadsheet users with systematic string manipulation~\cite{gulwani2011automating},
and help programmers use unfamiliar APIs~\cite{FengM0DR17,Neo,tfcoder};
\autoref{fig:pbe} shows example PBE tasks from three domains.

\paragraph{Challenge: Harnessing the Power of LLMs for PBE}

How can we automatically learn programs from the input-output examples like those shown in \autoref{fig:pbe}?
The traditional \emph{program synthesis} approach is based on combinatorial search~\cite{udupa2013transit,albarghouthi2013recursive,osera2015type,alur2018search,reynolds2019cvc},
which works well for small programs and restrictive DSLs,
but becomes infeasible as the program size and the DSL complexity grow.
At the other end of the spectrum, purely \emph{neural} approaches~\cite{devlin2017robustfill,wen2024grounding} use a neural model to predict the program from input-output examples;
unfortunately, even state-of-art LLMs like GPT-4o~\cite{openai_gpt4_2024}
struggle to predict an entire program in an unfamiliar DSL: 
when we asked GPT-4o to generate 10 programs for the running example above,
none of them were entirely correct.%
\footnote{A detailed analysis of GPT-4o's performance on this task is provided in \autoref{app:gpt4o-solutions}.}

In the past, the limitations of both program synthesis and neural techniques have motivated a hybrid approach,
where combinatorial search is \emph{guided} by a learned probabilistic model~\cite{balog2016deepcoder,kalyan2018neural,lee2018accelerating,odena2020bustle,tfcoder,shi2022crossbeam}.
Existing hybrid techniques, however, use domain-specific models
trained on datasets of similar PBE tasks,
which limits their generalization to new domains.
With the advent of LLMs, can we now use a single pre-trained model to guide program synthesis across a wide range of domains?

Interestingly, there is some tension in the hybrid approach
between the efficiency of the search algorithm and the power of the model:
a search algorithm is efficient when it \emph{factorizes the search space} (\ie, merges many search states into one),
which often makes it incompatible with a powerful model that requires a lot of context to make a prediction.
Specifically, one of the most widely used program synthesis techniques is \emph{bottom-up search}~\cite{albarghouthi2013recursive,udupa2013transit,barke2020just,tfcoder,arborist},
which is a dynamic programming algorithm,
whose efficiency relies on reusing the work of constructing and evaluating subprograms in many different contexts.
This essentially precludes using models with unlimited left-to-right context---like LLMs--to guide bottom-up search.
%

\paragraph{Our Solution: Context-Free LLM Approximation}

\begin{figure}[t]
\small
	\centering
	\includegraphics[width=.9\textwidth]{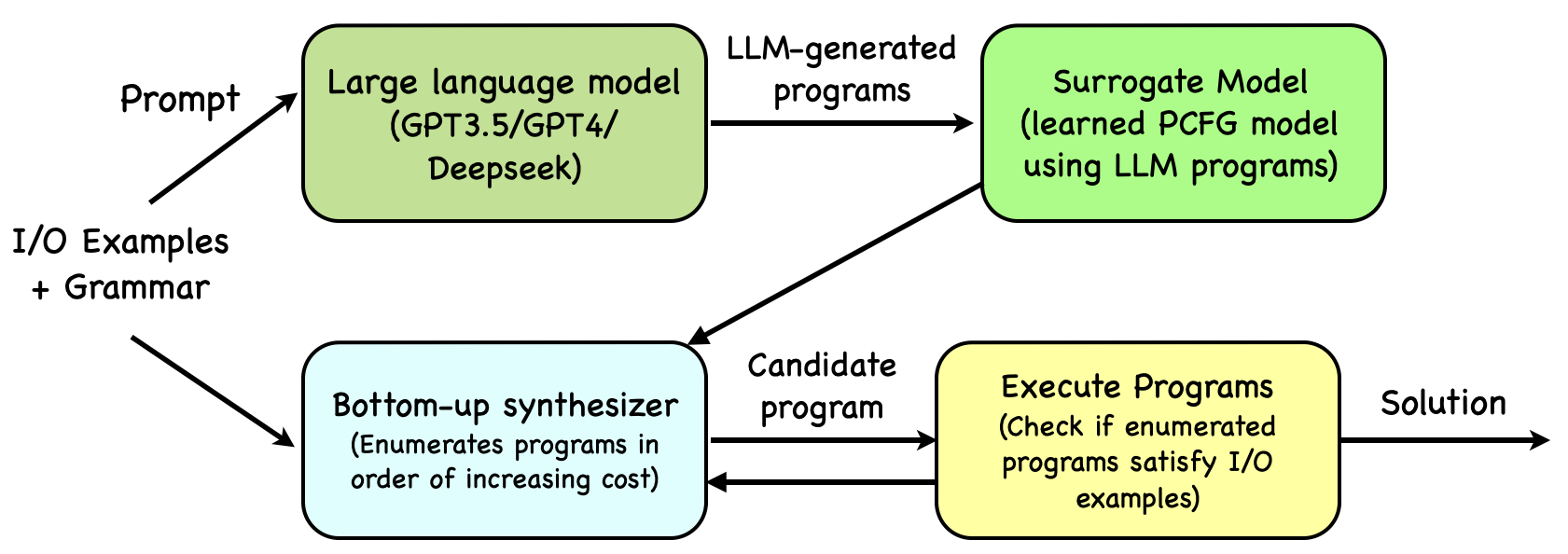}
	\caption{An overview of the hybrid program synthesis technique that uses a context-free LLM approximation.
	Programs generated by an LLM are used to learn a PCFG, which guides a bottom-up synthesizer to generate programs until a solution is found.
	}
	\label{fig:overview}
\end{figure}

To bridge this gap and harness the power of LLMs to guide bottom-up search,
we propose to approximate the LLM's conditional output distribution \emph{for a given task} with a context-free surrogate model.
Recent work in NLP~\cite{zhang2023tractable} has found that a Hidden Markov Model (HMM)
trained to match an LLM can be used as an efficient surrogate in style-controlled language generation.
We extend this idea to program synthesis,
replacing the HMM with a \emph{probabilistic context-free grammar} (PCFG).
The benefits of using a PCFG are twofold:
\begin{enumerate*}[label=(\arabic*)]
	\item PCFGs are context-free, which makes them compatible with bottom-up search for PBE~\cite{barke2020just,tfcoder}, and
	\item while a context-free model may make a poor approximation to an LLM's full joint, in a PBE setting it is able to reasonably approximate an LLM's conditional distribution over output programs \emph{for a given prompt}. 
\end{enumerate*}
%
The overview of our approach is shown in \autoref{fig:overview}.

\paragraph{Evaluation}

We implemented this technique in a tool \tool%
\footnote{The name stands for ``HYbrid SYNTHesis'' and is pronounced like the flower ``hyacinth''.}
and evaluated it on 299 PBE tasks from three domains:
\arc grid-based puzzles~\cite{chollet2019measure},
tensor manipulation tasks from \tfcoder~\cite{tfcoder},
and string manipulation tasks from the \sygus benchmark~\cite{alur2017sygus}, which are inspired by spreadsheet use cases.
Example problems from these domains are shown in \autoref{fig:pbe}.
Our evaluation shows that \tool outperforms both unguided search and LLMs alone, 
solving 58\% of the tasks overall, compared to 40\% for unguided search and 6\% for LLMs without search.
Our tool also outperforms baseline program synthesizers for these domains---\arga, \tfcoder, and \probe~\cite{barke2020just}, respectively;
importantly, in the \tensorbench domain, the guidance from the LLM not only speeds up the search,
but also frees the user from having to explicitly provide any non-standard \emph{constants} that the solution might use,
thereby significantly improving the usability of the tool.

\paragraph{Contributions}
In summary, this paper makes the following contributions:
\begin{enumerate}
	\item
	We propose a hybrid program synthesis approach that integrates LLMs with efficient bottom-up search
	via a task-specific context-free approximation.
	\item
	We implement this approach in a tool \tool and instantiate it on three domains: 
	grid-based puzzles (\arc), tensor manipulation (\tensorbench), and string manipulation (\stringbench).
	While the latter two domains reuse off-the-shelf bottom-up synthesizers,
	for \arc we implement a custom synthesizer that uses a divide-and-conquer strategy~\cite{alur2017scaling}
	to leverage the structure of the rule-based DSL to further speed up the search.
	\item
	We evaluate \tool on the three domains 
	and show that it outperforms both the LLM alone 
	and existing baseline synthesizers, which are not guided by LLMs.
\end{enumerate}
\section{Background}

\subsection{Programming-By-Example}

Programming by Example (PBE)~\cite{gulwani2016pbe} is the task of synthesizing programs that satisfy a given set of input-output examples.
To restrict the program space, the programs are typically drawn from a \emph{domain-specific language} (DSL),
which is specified by a \emph{context-free grammar} and an \emph{evaluation function}.
This section provides a formal definition of these concepts.


\paragraph{Context-Free Grammars}

A \emph{context-free grammar} (CFG) is a quadruple $\grammar = (\nontermset, \termset, \start, \ruleset)$, 
where $\nontermset$ is a set of non-terminal symbols, 
$\termset$ is a set of terminal symbols, 
$\start \in \nontermset$ denotes the starting non-terminal, 
and $\ruleset$ is the set of production rules.
%
%
An example CFG is shown in \autoref{fig:grammar}.
%
We denote with $\ruleset(\nterm)$ the set of all rules $\rl \in \ruleset$ whose left-hand side is $\nterm$.
%
%
A grammar $\grammar$ defines a (leftmost) \emph{single-step derivation} relation on sequences of symbols:
$s \nterm \alpha \step s \beta \alpha$ if $\nterm \to \beta \in \ruleset$, where $s \in \termset^*$ and $\alpha, \beta \in (\nontermset \cup \termset)^*$.
The transitive closure of this relation $\manystep$ is called (leftmost) \emph{derivation}.
%

\begin{figure}[t]
  \begin{minipage}{0.53\textwidth}
  \small    
  \begin{align*}
  \nonterm{Rule}  &\rightarrow && \T{if}\ \nonterm{Filter}\ \T{then}\ \nonterm{Transform}\\
  \nonterm{Filter} &\rightarrow && \nonterm{Atom} \mid \T{not}\ \nonterm{Atom}\mid \nonterm{Atom}\ \T{&&}\ \nonterm{Filter} \mid  \dots\\
  \nonterm{Atom} &\rightarrow && \nonterm{Color}\ \T{==}_c\ \nonterm{Color} \mid \nonterm{Size}\ \T{==}_s\ \nonterm{Size} \mid \dots\\
  \nonterm{Transform} &\rightarrow && \T{update\_color(}\nonterm{Color} \T{)} \mid \T{move(}\nonterm{Dir} \T{)} \mid \dots\\
  \end{align*}
  \end{minipage}%
  \begin{minipage}{0.47\textwidth}
  \small
  \begin{align*}
  \nonterm{Color} &\rightarrow && \T{color\_of(}\nonterm{Obj} \T{)} \mid \T{GREY} \mid \T{RED} \dots\\
  \nonterm{Size} &\rightarrow && \T{size\_of(}\nonterm{Obj} \T{)} \mid \T{MIN} \mid \T{MAX} \mid \dots\\
  \nonterm{Dir} &\rightarrow && \T{dir\_of(}\nonterm{Obj} \T{)} \mid \T{UP} \mid \T{DOWN} \mid \dots\\
  \nonterm{Obj} &\rightarrow && \T{self} \mid \T{x} \mid \T{y} \mid \dots\\
  \end{align*}
  \end{minipage}
  \caption{A fragment from the context-free grammar of our \arc DSL.}\label{fig:grammar}
\end{figure}

\paragraph{Programs}

A \emph{program} $\prog \in \Sigma^*$ is a terminal sequence derivable from some $\nterm \in \nontermset$;
we call a program \emph{whole} if it is derivable from $\start$.
The set of all programs is called the \emph{language} of the grammar $\grammar$:
$\lang(\grammar) = \{s \in \termset^* \mid \nterm \manystep s\}$.
The \emph{trace} of a program $\trace(\prog)$ is the sequence of production rules 
$\rl_1, \ldots, \rl_n$ used in its derivation ($\nterm \step \alpha_1 \step \ldots \step \alpha_{n-1} \step \prog$).
The \emph{size} of a program $|\prog|$ is the length of its trace.
The semantics of a program $\prog$ is defined by the evaluation function $\sem{\prog}\colon \vals^*\to \vals$,
which maps the values of program variables to its output value.

\paragraph{Problem Statement}

A PBE problem is defined by a DSL with a grammar $\grammar$ and an evaluation function $\sem{\cdot}$,
as well as a set of input-output examples $\examples = \many{\langle i, o\rangle}$ where $i\in\vals^*$, $o\in \vals$.
%
%
A \emph{solution} to the problem is a program $P \in \lang(\grammar)$ 
such that $\forall {\langle i, o\rangle} \in \examples$, $\sem{\prog}(i) = o$.

\subsection{Assigning Costs to Programs}\label{sec:costs}


\paragraph{Weighted Context-free Grammar}

A \emph{weighted context-free grammar} (WCFG) $\grammar_w$
is a pair of a CFG $\grammar$ and a function $\rwei: \ruleset \rightarrow \mathbb{R}^+$
that maps each production rule $\rl \in \ruleset$ to a positive weight.
Given a weighted grammar $\grammar_w$, we can define the \emph{real cost} of a program $\prog$ as the sum of weights of all the productions in its trace:
$\rcost(\prog) = \sum_{\rl_i\in \trace(\prog)} \rwei(\rl_i)$.

For the purposes of search, it is convenient to define a \emph{discrete weight} function $\wei: \ruleset \rightarrow \mathbb{Z}^+$,
which rounds weights up to the nearest integer:
$\wei(\rl) = \round{\rwei(\rl)}$.
The (discrete) \emph{cost} of a program $\prog$ is defined as the sum of discrete production weights:
$\costfn(\prog) = \sum_{\rl_i\in \trace(\prog)} \wei(\rl_i)$.
Note that because of error accumulation, the discrete cost of a program can differ from its rounded real cost,
but 
the difference can be made arbitrarily small by scaling all the costs by a constant factor $\alpha > 1$.

\paragraph{Probabilistic Context-free Grammar}

A popular way to assign weights to production rules is via a \emph{probabilistic context-free grammar} (PCFG).
A PCFG $\grammar_p$  is a pair of a CFG $\grammar$ and a function $\pcfg : \ruleset \rightarrow [0,1]$ 
that maps each production rule $\rl \in \ruleset$ to its probability, 
such that probabilities of all the rules for a given non-terminal $\nterm \in \nontermset$ sum up to one: 
$\forall \nterm .\sum_{\rl \in \ruleset(\nterm)} \pcfg(\rl) = 1$.
A PCFG defines a probability distribution on programs:
$\pcfg(\prog) = \prod_{\rl_i\in \trace(\prog)} \pcfg(\rl_i)$.

Given a PCFG $(\grammar, \pcfg)$ we can derive a WCFG $\grammar_w$ where $\rwei(\rl) = -\log(\pcfg(\rl))$;
to make sure that all weights are finite and positive,
we exclude rules with $\pcfg(\rl)=0$
and inline rules with $\pcfg(\rl)=1$.
In this WCFG, the real cost of a program is related to its probability: $\rcost(\prog) = -\log(\pcfg(\prog))$.

\subsection{Bottom-up Search}

\begin{algorithm}[t]
\small
\caption{Bottom-Up Search Algorithm}\label{algo}
\begin{algorithmic}[1]
\Require{Input-output examples $\examples$, a WCFG $\grammar_w = (\nontermset, \termset, \start, \ruleset, \wei)$}
\Ensure{A program $\prog$ consistent with $\examples$ or failure ($\bot$)}
\Procedure{Bottom-up-Search}{$\grammar_w, \examples$}
\State $\level, \bank, \exec \gets 1, \emptyset, \emptyset$ 
    \Comment{Initialize state of the search}
    \While{true}
      \For{$\prog \in \textproc{New-Programs}(\grammar_w, \level, \bank)$}  \Comment{For all programs of cost $\level$}
				
		    \State $\eval \gets [{\langle i, \sem{\prog}(i)\rangle} \mid \langle i, o\rangle \in \examples]$
            \Comment{Evaluate on inputs from $\examples$}	
        \If{($\eval = \examples$)}
          \State \textbf{return} $\prog$  \Comment{$\prog$ fully satisfies $\examples$, solution found!}
        \ElsIf{($\eval \in \exec$)}
		      \State{\textbf{continue}}                        \Comment{$\prog$ is semantically equivalent to another program in $\bank$}
        \EndIf    
        \State $\bank[\level] \gets \bank[\level] \cup \{\prog\}$ \Comment{Add to the bank, indexed by cost}
        \State $\exec \gets \exec \cup \eval$                                     \Comment{Cache evaluation result}
			\EndFor		
		\State $\level \gets \level + 1$
    \EndWhile
    \State \textbf{return} $\bot$    \Comment{Cost limit reached}
\EndProcedure
\Procedure{New-Programs}{$\grammar_w$, $\level$, $\bank$}
  \For{$\rl = \nterm \rightarrow s_0 \nterm_1 s_1 \nterm_2 \ldots \nterm_k s_k \in \ruleset$} \Comment{$\rl$ is a production rule with $k$ non-terminals}
      \For{$(c_1,\dots,c_k) \in \set{[1..\level-1]^k}{\sum c_i = \level - \wei(\rl)}$}        \Comment{For all subexpression costs}
      \For{$(P_1,\dots,P_k)\in \set{\bank[c_1]\times\ldots\times\bank[c_k]}{\bigwedge_i \nterm_i \manystep P_i}$} \Comment{For all subexpressions}
          \State	$\mathbf{yield}\ s_0 P_1 s_1 P_2 \dots\ P_k s_k$  \Comment{Substitute subexpressions into $\rl$'s RHS}
      \EndFor
      \EndFor
  \EndFor
\EndProcedure
\end{algorithmic}
\end{algorithm}

Bottom-up search is a popular search technique in program synthesis~\cite{albarghouthi2013recursive,udupa2013transit,barke2020just,tfcoder,arborist}, 
which enumerates programs from the DSL in the order of increasing costs
until it finds a program that satisfies the given examples.
%
%
The search is implemented as a dynamic programming algorithm (see \autoref{algo}),
which maintains a program \emph{bank} $\bank$ mapping discrete costs to programs of that cost.
Starting with an empty bank and current cost level $\level = 1$,
the search iteratively creates all programs of cost 1, 2, 3, and so on;
to create complex programs, the algorithm \emph{reuses} simpler programs already stored in the bank,
and combines them using the production rules of the grammar.

For example, consider the CFG in \autoref{fig:grammar},
and assume a uniform weight function $\wei(\cdot)=1$.
Then in the first iteration (cost level 1),
the algorithm will enumerate programs consisting of a single literal or variable---%
\eg \T{self}, \T{GREY}, \T{UP}, \etc---%
and store them in $\bank[1]$.
At cost level 2, it will enumerate unary operators applied to programs stored in $\bank[1]$:
\eg \T{color\_of(self)}, \T{move(UP)}, \etc.
More generally, at cost level $\level$,
the algorithm considers all available productions, 
and for each production, enumerates all combinations of arguments whose costs sum up to $\level - 1$.

%

During search, each candidate expression is evaluated to see if it satisfies the examples (lines 5--7).
Importantly, the search maintains a cache of all evaluation results $\exec$,
and discards the newly constructed program if it is \emph{observationally equivalent} to a program already in the bank (line 8),
\ie if it evaluates to the same output for all inputs in the examples.
This step is the key to the efficiency of the bottom-up search algorithm:
it allows the synthesizer to factorize the search space by evaluation result,
significantly reducing the number of programs explored at each cost level.

%

\section{The \tool Approach}\label{sec:approach}
A key challenge in program synthesis is the astronomical size of the search space the synthesizer has to explore.
For example, to find the program \autoref{eq:example-program}, the solution to the \arc task from the introduction,
bottom-up search with a uniform weight function has to enumerate around 450K 
programs (all programs of size $\leq 16$),
which takes 4.5 minutes in our experiments.

On the other hand, sampling solutions to this task from an LLM
yields programs that are \emph{close} to the desired solution, even if not quite correct.
As we show in \autoref{app:gpt4o-solutions},
GPT-4o uses relevant components \T{update\_color}, \T{color\_of}, and \T{is\_neighbor} in nearly all of its solutions
(usually missing some part of the filter or using the wrong color in the transform),
and never uses irrelevant components like \T{move} or \T{rotate}.
This suggests that the LLM generally has the right intuition about the components the solution needs to use;
our insight is to leverage this intuition to guide bottom-up search
by \emph{assigning lower weights to the components that the LLM uses frequently}.

\subsection{Guiding Bottom-up Search with Context-Free LLM Approximation}\label{sec:approach:general}
The overview of our approach, \tool, is shown in \autoref{fig:overview}.
Given a PBE problem consisting of a DSL with grammar $\grammar$ and a set of input-output examples $\examples$,
\tool proceeds in three steps.

\paragraph{Step 1: Sampling Solutions from an LLM}

\tool starts by creating an LLM prompt that contains $\grammar$ and $\examples$;
the prompt can be optionally augmented with in-context examples if they are available for the given DSL.
A complete prompt for the \arc running example can be found in \autoref{app:llm-prompt}.
The LLM is then used to sample a set $\{S_i\}_{i=1}^{\nsol}$ of completions;
the choice of $\nsol$ trades off computational cost and
the faithfulness of the approximation to the true LLM conditional.

\paragraph{Step 2: Learning a PCFG from LLM Solutions}

Next, \tool attempts to parse each completion $S_i$ into a program $P_i$ using the grammar $\grammar$.
%
The resulting set of programs $\{P_i\}_{i=1}^{\nsol'}$ (where $\nsol' \leq \nsol$) is used to learn a PCFG $\grammar_p$ via maximum likelihood estimation:
$\pcfg(\rl) = \frac{\mathrm{count}(\rl) + \alpha}{\sum_{\rl \in \ruleset} \mathrm{count}(\rl) + \alpha \times |\ruleset|}$.
%
Here $\mathrm{count}(\rl)$ is the frequency of rule $\rl$ in all the derivations of the programs in $\{P_i\}$ and
$\alpha$ is a smoothing parameter that ensures that every rule has a non-zero probability (typically set to 1).

Our experiments show that some models struggle to generate grammatical completions, leading to $\nsol' \ll \nsol$.
To increase the sampling efficiency in those cases,
\tool implements \emph{non-strict mode}, where ungrammatical completions $S_i$ are not discarded.
Instead the tool performs lexical analysis on $S_i$ to convert it into a sequence of terminals 
and approximates the frequency of each production $\rl$
based on the frequency of its \emph{operator terminal},
a designated terminal of $\rl$, which represents a DSL operator;
\eg $\mathrm{count}(\nonterm{Atom} \to \T{not}\ \nonterm{Atom}) = \mathrm{count}(\T{not})$.%
\footnote{Typically, the operator terminal uniquely identifies $\rl$,
but when this is not the case, we can normalize $\mathrm{count}(\rl)$ by the number of rules in $\ruleset$ that produce this terminal.}



\paragraph{Step 3: Guiding Bottom-up Search with PCFG}

Finally, \tool uses the PCFG  computed in the previous step
to derive a weighted grammar $\grammar_w$ as explained in \autoref{sec:costs},
and uses it to initialize the bottom-up search procedure in \autoref{algo}.
As a result, the search is guided by the insights from the the LLM.
For example, the WCFG learned from the GPT-4o completions for the \arc task above
gives the relevant transform operator \T{update\_color} weight 2,
while all other $\nonterm{Transform}$ rules have weight 4;
the relevant filter operators \T{color\_of} and \T{is\_neighbor} are similarly down-weighted.
As a result, the search procedure only has to enumerate around 220K programs instead of 450K,
achieving a 4x speedup,
and solving the motivating example in just one minute with LLM guidance.

\subsection{Domain-Specific Instantiations}

We now describe how the \tool approach is instantiated in three different domains: 
\arc grid puzzles, \tensorbench manipulations, and \stringbench manipulations.

\paragraph{\arc Domain}

An example task from this domain is shown in \autoref{fig:pbe-a}
and has been used as a running example throughout this paper.
There is no established DSL for \arc,
and arguably, DSL design is the biggest challenge when attempting to solve \arc using a PBE approach,
since it is hard to capture the wide variety of tasks in this domain.
Our DSL is inspired by the rule-based language of \arga~\cite{xu2023graphs},
which we modified slightly to make it more compositional.

A program in our DSL is a sequence of rules of the form \T{if} \emph{filter} \T{then} \emph{transform}.
A rule refers to the current object \T{self},
which is modified by the transform if the filter is satisfied in the current state of the grid.
The rule can also refer to other objects in the grid, such as \T{other} in \autoref{eq:example-program}.
This program is well-defined because its filter uniquely identifies the object \T{other};
if the filter is too weak to uniquely determine the effect of the transform, 
the program's output is considered undefined.
%
%
The full grammar of our DSL can be found in \autoref{app:arc-grammar}.

Instead of searching for a complete program using \autoref{algo},
we further optimize our synthesizer using a divide-and-conquer strategy inspired by~\cite{alur2017scaling},
searching for filters and transforms \emph{separately}.
Specifically, \tool-\arc first searches for transforms that are correct on some objects in the grid;
once it has found a set of transforms that collectively describe all grid objects,
it searches for filters that distinguish between the subsets of objects changed by each transform.

Consider once again our running example.
When the transform synthesizer enumerates the expression \T{update\_color(color\_of(other))},
it detects that this transform works for all \emph{grey objects}, 
because for each grey object \T{self} there exists a corresponding object \T{other} whose color can be copied.
Now the goal of filter synthesis is to find a boolean expression that holds
exactly for those pairs of objects $(\T{self}, \T{other})$ that make the transform work.
See \autoref{app:arc-algo} for more details about this algorithm.

\paragraph{\tensorbench Domain}

This domain originates from the \tfcoder synthesizer~\cite{tfcoder},
which takes as input examples of a tensor transformation (with an optional natural language description)
and synthesizes a \tensorflow program that performs the transformation. 
%
An example task from this domain is shown in \autoref{fig:pbe-b},
whose solution is: 
\T{tf.gather_nd(in1, tf.stack((in2, in3), axis=-1))}.
%
%
The main challenge, however, is that the \tensorflow grammar is very large (see \autoref{sec:tensor-grammar}),
and most importantly, the programs are allowed to use an \emph{unbounded} set of constants.
The original \tfcoder synthesizer requires the user to provide any non-standard constants that a task might require,
and, according to their paper,
this is the main barrier to the usability of their tool.

For program synthesis in this domain we use the \tfcoder synthesizer off the shelf.
\tfcoder performs weighted bottom-up search, 
using a combination of hand-tuned weights and weights derived by two custom-trained neural models.
\tool-\tensorbench replaces these weights entirely with weights computed by sampling from an LLM.
Importantly, our version of the tool does not require the user to provide any constants;
instead we extract constants from the LLM completions,
whereby significantly reducing the burden on the user.

\paragraph{\stringbench Domain}

Our third domain involves string manipulation tasks from the \sygus competition~\cite{AlurBJMRSSSTU13},
which are inspired by spreadsheet use cases.
An example task, which requires extracting the top-level domain name from a URL, 
is shown in \autoref{fig:pbe-c}.
%
%
In this domain we use the \probe~\cite{barke2020just} synthesizer off the shelf.
\probe performs weighted bottom-up search, 
starting with a uniform grammar and updating the weights on the fly;
\tool-\stringbench instead initializes \probe's search with weights derived from an LLM,
and disables the weight updates during search.
\vspace{-7pt}
\section{Experiments and Results}
\subsection{Experimental Setup}
We evaluate \tool on 299 PBE tasks from three different domains: \arc (160 tasks), \stringbench (70 tasks) and \tensorbench (69 tasks).
\paragraph{\arc Benchmark}
The 160 \arc tasks are taken from the testing set of \arga \cite{xu2023graphs}.
This \emph{object-centric} subset of the full \arc corpus is known as \ocarc,
and has been used to evaluate other \arc solvers~\cite{lei2024generalized}.
\arc specifications consist of 2-7 input-output training grids and 1 testing grid.
Correctness is based on whether the generated solution produces the correct output on the testing grid.
Our \arc DSL has a total of 20 operations and 50 constants and variables across all types.

\paragraph{\tensorbench Benchmark}
The 69 \tensorbench tasks taken from \tfcoder focus on tensor manipulation. 49 of them are sourced from StackOverflow inquiries, and 20 are from real-world scenarios faced by \tensorflow users at Google.
%
%
The overall benchmark suite consists of 72 tasks. We use three of these tasks as in-context examples and evaluate on the rest.
The grammar for this domain consists of 134 Tensorflow operations, primitives like \scode{0, 1, -1, True} and other task-specific constants.

\paragraph{\stringbench Benchmark}
The 70 \stringbench tasks are taken from testing set of \probe, which is derived from the \sygus benchmark~\cite{AlurBJMRSSSTU13}.
%
%
The number of examples ranges from 2 to 400.
The original \sygus benchmark have custom grammars for each task,
but we use a union of all the grammars to make the search more challenging;
the union grammar has 16 operations and 59 constants.

\paragraph{Configurations}
Our main \tool configuration uses \gptfouro as the LLM,
with 100 samples per task to learn a PCFG in non-strict mode
(\ie syntactically invalid completions are included in the PCFG learning process, as explained in \autoref{sec:approach:general}).
For each domain, we compare the performance of \tool with
a baseline synthesizer for that domain (\arga%
\footnote{At the time of writing, \arga is no longer state of the art on the \ocarc dataset;
we explain in \autoref{sec:related} why the comparison with \arga is still relevant.}, \probe, and \tfcoder),
as well as three ablations:
\begin{enumerate*}[label=(\arabic*)]
	\item \emph{no search}, \ie using the 100 samples from the LLM directly,
	\item \emph{unguided search}, \ie running the same synthesizer but with a uniform weighted grammar,
 and \item \emph{binary surrogate}, running the synthesizer but with a \emph{binary PCFG},
 \ie a CFG that includes the components present in the LLM samples with equal probabilities,
 and excludes all other components completely.
\end{enumerate*}
We also analyze the performance of \tool with different number of samples used to learn the PCFG (10, 20, and 50),
with other LLMs (\gptthree and \deepseek~\cite{guo2024deepseek}),
as well as in strict mode (which discards syntactically invalid LLM completions).
The timeout is set to 10 minutes for all experiments and includes the search time and time to sample LLM completions (and compute PCFG).
The average time to sample 100 solutions from \gptfouro is 4 seconds, 12 seconds and 20 seconds per task for the \stringbench, \arc and \tensorbench domains, respectively.
\subsection{Results}
\begin{figure}[t]
\begin{subfigure}{.49\textwidth}
\centering
\includegraphics[width=0.92\textwidth]{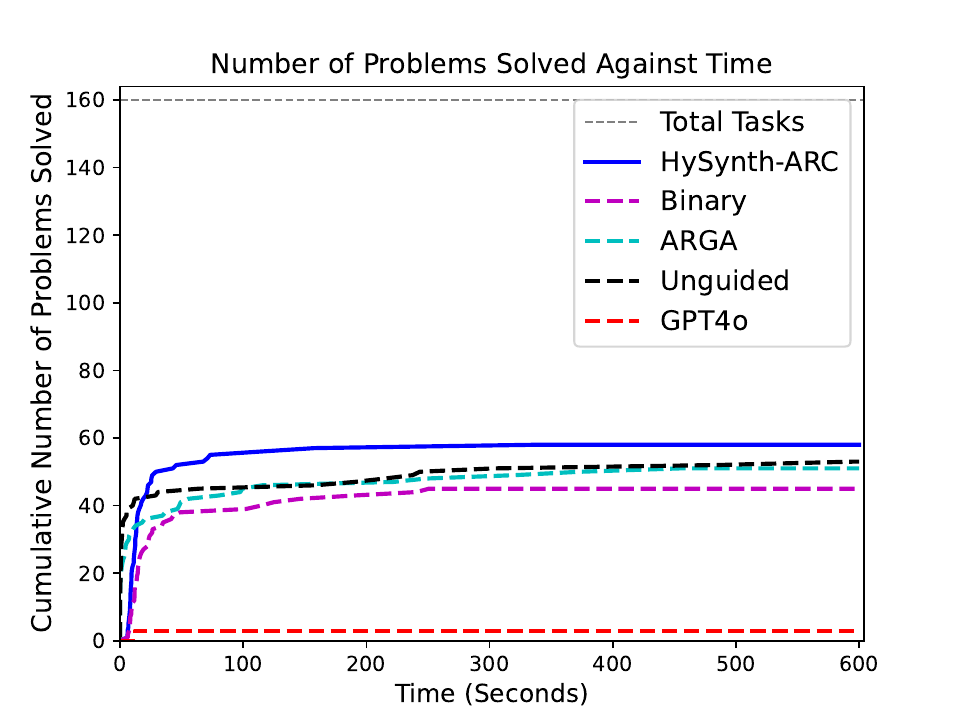}
\caption{\tool-\arc results with \gptfouro}
\label{fig:arc-gpt4o}
\end{subfigure}%
\begin{subfigure}{.49\textwidth}
\centering
\includegraphics[width=0.92\textwidth]{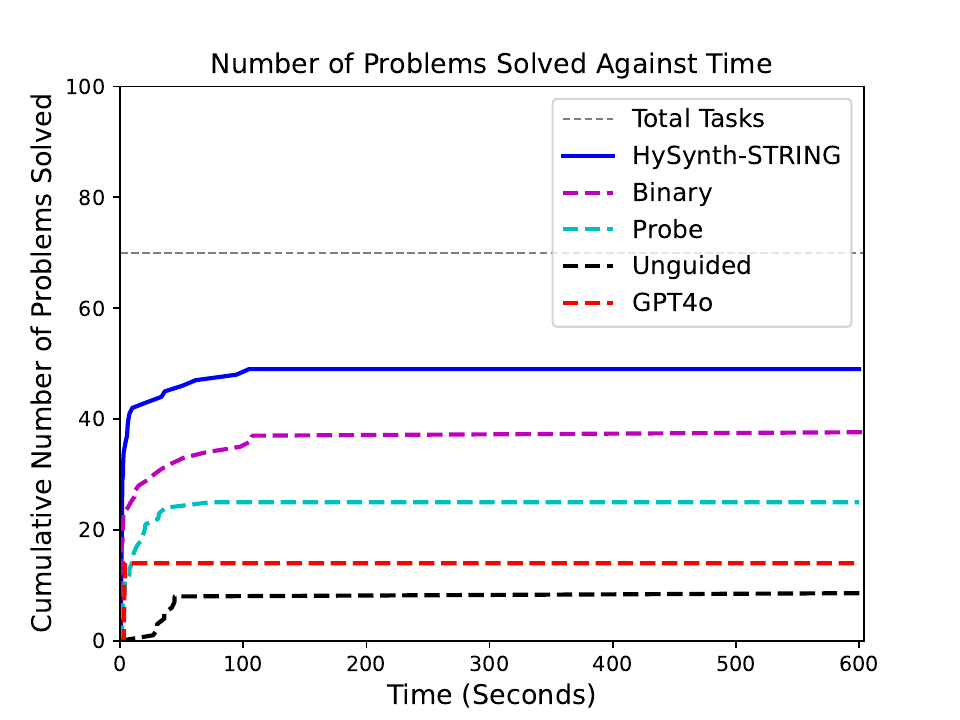}
\caption{\tool-\stringbench results with \gptfouro}
\label{fig:string-gpt4o}
\end{subfigure}%

\begin{subfigure}{.49\textwidth}
\centering
\includegraphics[width=0.92\textwidth]{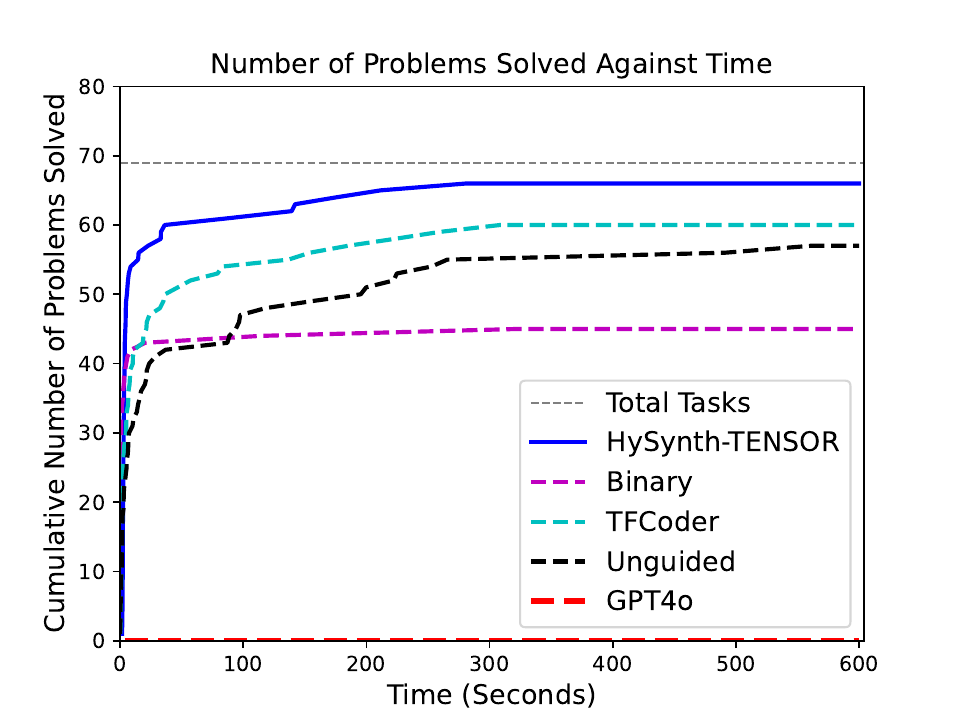}
\caption{\tool-\tensorbench results with \gptfouro} \label{fig:tfcoder-gpt4o}
\end{subfigure}%
\begin{subfigure}{.49\textwidth}
\centering
\resizebox{0.95\textwidth}{!}{
\renewcommand{\arraystretch}{1.5} 
\begin{tabular}{|c|c|c|}
	\hline
	\textbf{Domain/Model} & \textbf{\% Valid completions} \\
	\hline
	\tensorbench-\gptfouro & 99.9\% \\
	\hline
	\tensorbench-\deepseek & 92.8\% \\
	\hline
	\stringbench-\gptfouro & 37.5\% \\
	\hline
	\stringbench-\deepseek & 0\% \\
	\hline
	\arc-\gptfouro & 78.4\% \\
	\hline
\end{tabular}
}
\caption{Percentage of syntactically valid completions}
\label{fig:validity_totals}
\end{subfigure}%
\caption{(a,b,c) Number of benchmarks solved by \tool as a function of time for the \arc, \tensorbench, and \stringbench domains; timeout is 10 min. (d) Percentage of syntactically valid completions per domain.}\label{fig:eval}
\end{figure}

\paragraph{How does \tool compare to baselines and ablations?}
We compare the time to solution for the main \tool configuration, baseline synthesizers, and the three ablations;
the results for the three domains are shown in \autoref{fig:arc-gpt4o}, \autoref{fig:string-gpt4o}, and \autoref{fig:tfcoder-gpt4o}.
Overall, \tool consistently outperforms both the baseline synthesizers and ablations,
solving more tasks across all domains. 

In more detail, direct LLM sampling performs very poorly on all domains, solving between 0 and 14 tasks;
this confirms our hypothesis that LLMs struggle on PBE tasks in domain-specific languages,
which are not prevalent in their training data.
Interestingly, despite not being able to solve \emph{any} \tensorbench tasks by itself,
\gptfouro provides excellent guidance for \tool on that domain,
helping it solve 96\% of the total benchmark!
On the other hand, synthesis guided by a binary surrogate model performs worse than \tool (and even unguided search in case of \arc and \tensorbench) since the search excludes essential components from the grammar.

\begin{figure}[htp]
\centering
\begin{subfigure}{.49\textwidth}
\centering
\includegraphics[width=0.92\textwidth]{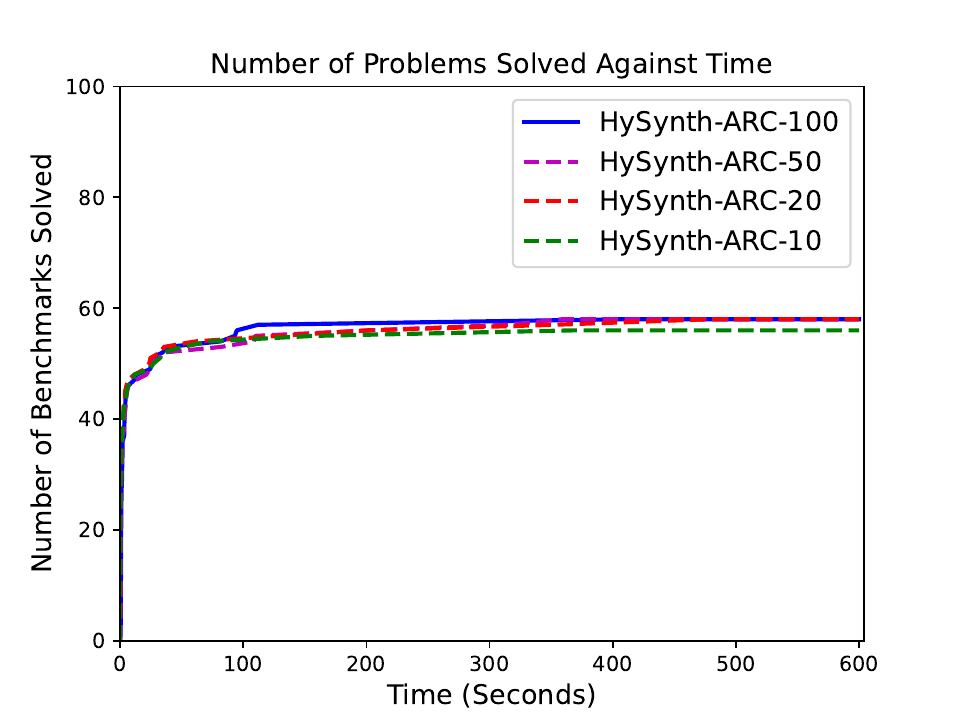}
\caption{\tool-\arc with varied sample sizes}
\label{fig:arc-samples}
\end{subfigure}%
\begin{subfigure}{.49\textwidth}
\centering
\includegraphics[width=0.92\textwidth]{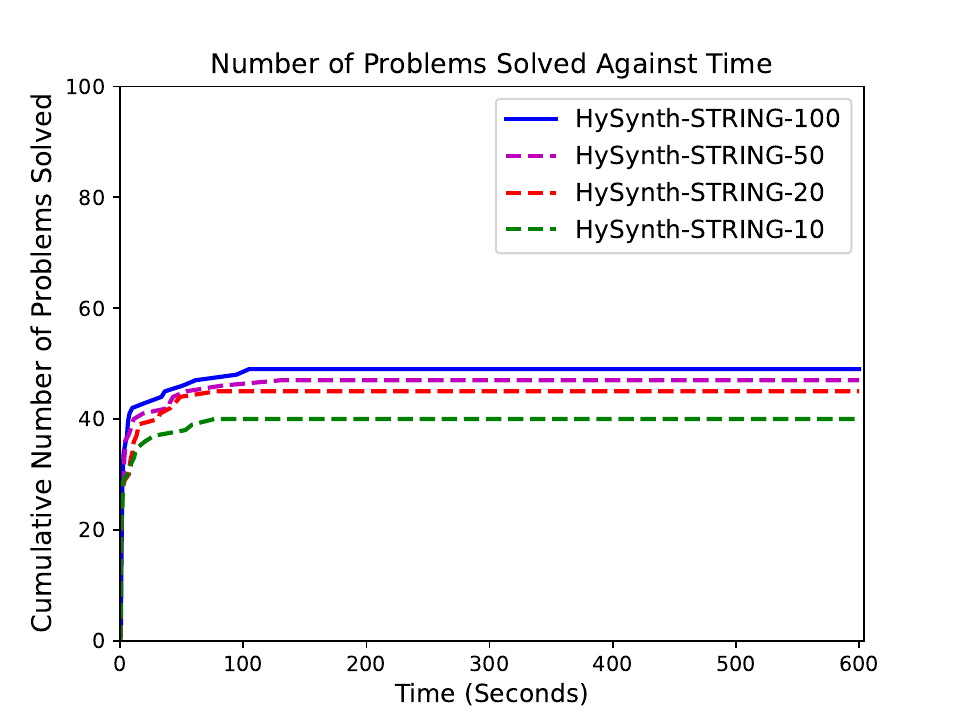}
\caption{\tool-\stringbench with varied sample sizes}
\label{fig:string-samples}
\end{subfigure}%

\begin{subfigure}{.49\textwidth}
\centering
\includegraphics[width=0.92\textwidth]
{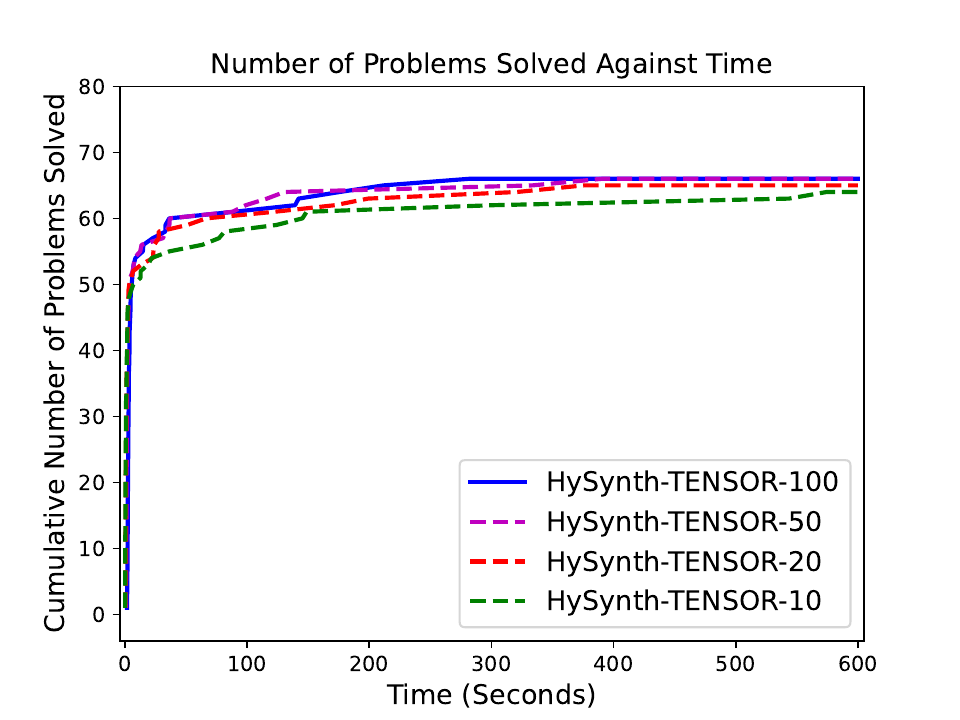}\hfill
\caption{\tool-\tensorbench with varied sample sizes}
\label{fig:tf-samples}
\end{subfigure}%
\begin{subfigure}{.49\textwidth}
\includegraphics[width=0.93\textwidth]{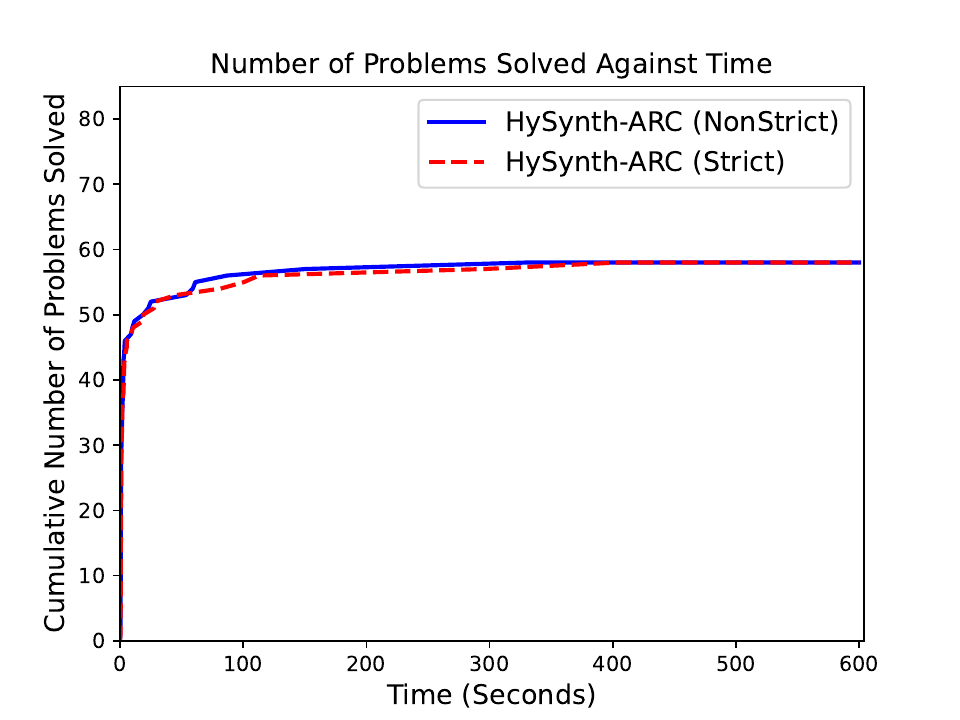}
\caption{\tool-\arc with strict and non-strict modes}
\label{fig:strict}
\end{subfigure}%
\caption{\tool-\arc, \tool-\tensorbench and \tool-\stringbench results guided by a PCFG learned from different number of \gptfouro samples (n=10, 20, 50, 100).}
\label{fig:pcfg-samples}
\end{figure}

In \stringbench and \tensorbench domains, the baseline synthesizers predictably do better than unguided search,
since both use the same search implementation, but with different weights.
On \arc, however, our custom synthesizer outperforms \arga%
\footnote{\cite{xu2023graphs} report 57 tasks for \arga but we could only reproduce 51 on our hardware with a 10 minute timeout.}
even without LLM guidance;
this speaks to the efficiency of the bottom-up search and the divide-and-conquer strategy we use,
which are results of years of research in the program synthesis community.
\paragraph{How many samples are needed to learn a PCFG?}
To better understand how the number of samples affects the quality of PCFG guidance,
we vary the number of \gptfouro programs used in PCFG learning $\nsol = 10, 20, 50, 100$,
and once again measure the number of tasks solved over time.
The results are shown in \autoref{fig:arc-samples}, \autoref{fig:string-samples}, and \autoref{fig:tf-samples}.
As expected, larger sample sizes generally lead to better performance, but the difference is minimal:
in \arc and \tensorbench, the difference between the best and worst performing versions of \tool is only 2 problems each,
while in \stringbench, \tool solves 9 fewer problems with 10 samples than with 100.
Despite these differences, all versions of \tool still outperform the baseline and unguided search.
This suggests that fewer samples are sufficient to effectively train a robust surrogate model, thereby optimizing costs.
\paragraph{Do our results generalize to other models?}

To answer this question, we repeat our experiments on \stringbench and \tensorbench domains
with \gptthree and the open-source model \T{deepseek-coder-33b-instruct} (\deepseek)~\cite{guo2024deepseek}.
The results with these models are detailed in \autoref{fig:other-models} in \autoref{app:other-models},
and they corroborate the pattern observed with \gptfouro,
where the guided versions outperform the baseline, unguided search, and direct sampling from the LLM.

\paragraph{How important is non-strict mode?}

\autoref{fig:validity_totals} shows the percentage of syntactically valid completions
generated by \gptfouro and \deepseek (where applicable).
You can see that while on \tensorbench almost all completions are valid,
this percentage falls to 78.4\% for \arc and 37.5\% for \stringbench;
this is not surprising, given that the former are \tensorflow programs, which the model has seen during training,
while the latter two are custom DSLs.
%
%
In the \stringbench benchmark, the grammar is very restricted (\eg only numeric constants allowed are 0-9), and the LLM has trouble adhering to this restricted grammar.
But even if we were to relax the definition of syntactic validity, LLM solutions would achieve a syntactic validity of only 47\%.
Hence our non-strict mode proves especially helpful for low-resource domains,
where otherwise we would have to discard a large proportion of completions.
At the same time, we find that \emph{given the same number of completions to learn from},
the PCFGs learned in non-strict mode are just as effective as those learned in strict mode:
as shown in \autoref{fig:strict}, \tool-\arc with the guidance from 100 \gptfouro completions solves 58 tasks \emph{in either mode}
(with the difference that strict mode has to sample more completions to get 100 valid ones).

\subsection{Limitations}

The main limitation of our hybrid approach \wrt purely neural approaches is that it requires implementing a synthesizer for each DSL of interest;
although we have shown that the same bottom-up search can be used across different domains,
some implementation effort is still required.
On the other hand, compared to purely symbolic approaches,
our method requires sampling from an LLM, which is costly;
additionally, the guidance provided by our approach is only as good as the LLM's completions:
if they contain many irrelevant operators,
our guided search can be \emph{slower} than unguided search.
Finally, our experiments are subject to the usual threat that the LLMs might have seen our benchmarks in their training data;
we do not consider it a major issue, however,
given that our main result is the superior performance of guided search \emph{relative} to using LLMs without search.
\section{Related Work}\label{sec:related}

\paragraph{Guiding Program Synthesis with Probabilistic Models}

The traditional approach to \emph{program synthesis} is based on combinatorial search~\cite{alur2018search},
augmented with pruning techniques based on program semantics~\cite{udupa2013transit,albarghouthi2013recursive,alur2017scaling}.
To further speed up the search, researchers have proposed \emph{guiding} the search with a learned probabilistic model.
%
Most approaches to guided search use special-purpose models that have to be trained on a domain-specific corpus of programs~\cite{lee2018accelerating}
or PBE tasks~\cite{balog2016deepcoder,kalyan2018neural,odena2020bustle,shi2022crossbeam}.
Although some of these models can be trained on synthetic data,
the training process is still expensive and requires manual tuning,
which makes it hard to apply these techniques to new domains.
%

With the advent of pretrained Large Language Models (LLMs), 
it seems only natural to use them to guide search-based program synthesis,
thus alleviating the need for domain-specific training data.
We are only aware of one other attempt to do this: 
concurrent work by \citet{li2024guiding},
which also extracts a PCFG from the LLM's samples, similarly to \tool.
%
An important difference is that they use the PCFG to guide \emph{top-down} \astar search, 
while we use it to guide \emph{bottom-up} search, which is known to be more efficient
(they also evaluate their tool on synthesis from logical formulas as opposed to PBE).

\paragraph{Solving the Abstraction and Reasoning Corpus}
%
All state-of-the-art solvers for this benchmark have relied on carefully curated DSLs for \arc \cite{butt2023codeit, arc_kaggle_first, alford2022neural, lei2024generalized, fischer2020solving}.
\citet{xu2023graphs} proposed the DSL we extend in our approach, and the \ocarc subset we evaluate on.
\citet{lei2024generalized} embed their DSL as a subset of PDDL and use a Generalized Planning (GP) algorithm as their search component.
They have the current best performance on \ocarc,  however they encode more domain-knowledge in the form of preconditions and per-abstraction restrictions on filters and transforms, to make GP viable.
Our approach does not require this additional information.
\cite{alford2022neural, banburski2020dreaming} use DreamCoder \cite{ellis2020dreamcoder}, to perform execution-guided search over a DSL for grid manipulations, however they only provide proof-of-concept evaluations.
\cite{wang2023hypothesis, tan2023large} also use an LLM to generate code given the spec of the task.
Both of these approaches interact with the model across several rounds, while our technique uses the suggestions from the LLM only as a starting point.
Our technique also performs a complete search guided by the LLM distribution, enabled by the structure of our DSL, whereas previous approaches only consider code directly generated by the LLM.
\section{Conclusion and Future Work}

Our approach introduces a robust technique for using both valid and invalid completions from an LLM to learn a surrogate model. By incorporating ungrammatical completions, we can extract useful insights that would otherwise be discarded. Overall, we provide an alternative to the conventional strategy of large-scale sampling from LLMs, proposing a more effective use of the available completions to guide the search process.
An interesting future direction would be to guide search with a more expressive context-dependent surrogate model. 

\bibliography{acl_natbib}

\begin{thebibliography}{46}
\expandafter\ifx\csname natexlab\endcsname\relax\def\natexlab#1{#1}\fi

\bibitem[{arc(2020)}]{arc_kaggle}
 2020.
\newblock \href {https://www.kaggle.com/competitions/abstraction-and-reasoning-challenge/leaderboard} {Arc kaggle competition leaderboard}.
\newblock Accessed: 2024-05-19.

\bibitem[{Albarghouthi et~al.(2013)Albarghouthi, Gulwani, and Kincaid}]{albarghouthi2013recursive}
Aws Albarghouthi, Sumit Gulwani, and Zachary Kincaid. 2013.
\newblock Recursive program synthesis.
\newblock In \emph{International Conference on Computer Aided Verification}, pages 934--950. Springer.

\bibitem[{Alford et~al.(2022)Alford, Gandhi, Rangamani, Banburski, Wang, Dandekar, Chin, Poggio, and Chin}]{alford2022neural}
Simon Alford, Anshula Gandhi, Akshay Rangamani, Andrzej Banburski, Tony Wang, Sylee Dandekar, John Chin, Tomaso Poggio, and Peter Chin. 2022.
\newblock Neural-guided, bidirectional program search for abstraction and reasoning.
\newblock In \emph{Complex Networks \& Their Applications X: Volume 1, Proceedings of the Tenth International Conference on Complex Networks and Their Applications COMPLEX NETWORKS 2021 10}, pages 657--668. Springer.

\bibitem[{Alur et~al.(2013)Alur, Bod{\'{\i}}k, Juniwal, Martin, Raghothaman, Seshia, Singh, Solar{-}Lezama, Torlak, and Udupa}]{AlurBJMRSSSTU13}
Rajeev Alur, Rastislav Bod{\'{\i}}k, Garvit Juniwal, Milo M.~K. Martin, Mukund Raghothaman, Sanjit~A. Seshia, Rishabh Singh, Armando Solar{-}Lezama, Emina Torlak, and Abhishek Udupa. 2013.
\newblock Syntax-guided synthesis.
\newblock In \emph{{Formal Methods in Computer-Aided Design, {FMCAD} 2013}}, pages 1--8.

\bibitem[{Alur et~al.(2017{\natexlab{a}})Alur, Fisman, Singh, and Solar-Lezama}]{alur2017sygus}
Rajeev Alur, Dana Fisman, Rishabh Singh, and Armando Solar-Lezama. 2017{\natexlab{a}}.
\newblock Sygus-comp 2017: Results and analysis.
\newblock \emph{arXiv preprint arXiv:1711.11438}.

\bibitem[{Alur et~al.(2017{\natexlab{b}})Alur, Radhakrishna, and Udupa}]{alur2017scaling}
Rajeev Alur, Arjun Radhakrishna, and Abhishek Udupa. 2017{\natexlab{b}}.
\newblock Scaling enumerative program synthesis via divide and conquer.
\newblock In \emph{{International Conference on Tools and Algorithms for the Construction and Analysis of Systems}}, pages 319--336. Springer.

\bibitem[{Alur et~al.(2018)Alur, Singh, Fisman, and Solar-Lezama}]{alur2018search}
Rajeev Alur, Rishabh Singh, Dana Fisman, and Armando Solar-Lezama. 2018.
\newblock Search-based program synthesis.
\newblock \emph{Communications of the ACM}, 61(12):84--93.

\bibitem[{Bai et~al.(2023)Bai, Wu, Chen, Wang, and Zhang}]{bai2023constituency}
Xuefeng Bai, Jialong Wu, Yulong Chen, Zhongqing Wang, and Yue Zhang. 2023.
\newblock Constituency parsing using llms.
\newblock \emph{arXiv preprint arXiv:2310.19462}.

\bibitem[{Balog et~al.(2016)Balog, Gaunt, Brockschmidt, Nowozin, and Tarlow}]{balog2016deepcoder}
Matej Balog, Alexander~L Gaunt, Marc Brockschmidt, Sebastian Nowozin, and Daniel Tarlow. 2016.
\newblock Deepcoder: Learning to write programs.
\newblock \emph{arXiv preprint arXiv:1611.01989}.

\bibitem[{Banburski et~al.(2020)Banburski, Gandhi, Alford, Dandekar, Chin, and tomaso~a poggio}]{banburski2020dreaming}
Andrzej Banburski, Anshula Gandhi, Simon Alford, Sylee Dandekar, Sang Chin, and tomaso~a poggio. 2020.
\newblock \href {https://openreview.net/forum?id=-gjy2V1ko6t} {Dreaming with {ARC}}.
\newblock In \emph{Learning Meets Combinatorial Algorithms at NeurIPS2020}.

\bibitem[{Barke et~al.(2020)Barke, Peleg, and Polikarpova}]{barke2020just}
Shraddha Barke, Hila Peleg, and Nadia Polikarpova. 2020.
\newblock Just-in-time learning for bottom-up enumerative synthesis.
\newblock \emph{Proceedings of the ACM on Programming Languages}, 4(OOPSLA):1--29.

\bibitem[{Berglund et~al.(2023)Berglund, Tong, Kaufmann, Balesni, Stickland, Korbak, and Evans}]{berglund2023reversal}
Lukas Berglund, Meg Tong, Max Kaufmann, Mikita Balesni, Asa~Cooper Stickland, Tomasz Korbak, and Owain Evans. 2023.
\newblock The reversal curse: Llms trained on" a is b" fail to learn" b is a".
\newblock \emph{arXiv preprint arXiv:2309.12288}.

\bibitem[{Butt et~al.(2023)Butt, Manczak, Wiggers, Rainone, Zhang, Defferrard, and Cohen}]{butt2023codeit}
Natasha Butt, Blazej Manczak, Auke Wiggers, Corrado Rainone, David~W Zhang, Micha{\"e}l Defferrard, and Taco Cohen. 2023.
\newblock Codeit: Abstract reasoning with iterative policy-guided program synthesis.

\bibitem[{Chollet(2019)}]{chollet2019measure}
Fran{\c{c}}ois Chollet. 2019.
\newblock On the measure of intelligence.
\newblock \emph{arXiv preprint arXiv:1911.01547}.

\bibitem[{Devlin et~al.(2017)Devlin, Uesato, Bhupatiraju, Singh, Mohamed, and Kohli}]{devlin2017robustfill}
Jacob Devlin, Jonathan Uesato, Surya Bhupatiraju, Rishabh Singh, Abdel-rahman Mohamed, and Pushmeet Kohli. 2017.
\newblock Robustfill: Neural program learning under noisy i/o.
\newblock In \emph{International conference on machine learning}, pages 990--998. PMLR.

\bibitem[{Ellis et~al.(2020)Ellis, Wong, Nye, Sable-Meyer, Cary, Morales, Hewitt, Solar-Lezama, and Tenenbaum}]{ellis2020dreamcoder}
Kevin Ellis, Catherine Wong, Maxwell Nye, Mathias Sable-Meyer, Luc Cary, Lucas Morales, Luke Hewitt, Armando Solar-Lezama, and Joshua~B Tenenbaum. 2020.
\newblock Dreamcoder: Growing generalizable, interpretable knowledge with wake-sleep bayesian program learning.
\newblock \emph{arXiv preprint arXiv:2006.08381}.

\bibitem[{Feng et~al.(2018)Feng, Martins, Bastani, and Dillig}]{Neo}
Yu~Feng, Ruben Martins, Osbert Bastani, and Isil Dillig. 2018.
\newblock \href {https://doi.org/10.1145/3192366.3192382} {Program synthesis using conflict-driven learning}.
\newblock In \emph{Proceedings of the 39th ACM SIGPLAN Conference on Programming Language Design and Implementation}, PLDI 2018, pages 420--435, New York, NY, USA. Association for Computing Machinery.

\bibitem[{Feng et~al.(2017)Feng, Martins, Wang, Dillig, and Reps}]{FengM0DR17}
Yu~Feng, Ruben Martins, Yuepeng Wang, Isil Dillig, and Thomas~W. Reps. 2017.
\newblock Component-based synthesis for complex apis.
\newblock In \emph{POPL}.

\bibitem[{Fischer et~al.(2020)Fischer, Jakobs, M{\"u}cke, and Morik}]{fischer2020solving}
Raphael Fischer, Matthias Jakobs, Sascha M{\"u}cke, and Katharina Morik. 2020.
\newblock Solving abstract reasoning tasks with grammatical evolution.
\newblock In \emph{LWDA}, pages 6--10.

\bibitem[{Gulwani(2011)}]{gulwani2011automating}
Sumit Gulwani. 2011.
\newblock Automating string processing in spreadsheets using input-output examples.
\newblock \emph{ACM Sigplan Notices}, 46(1):317--330.

\bibitem[{Gulwani(2016)}]{gulwani2016pbe}
Sumit Gulwani. 2016.
\newblock Programming by examples (and its applications in data wrangling).
\newblock In Javier Esparza, Orna Grumberg, and Salomon Sickert, editors, \emph{Verification and Synthesis of Correct and Secure Systems}. {IOS} Press.

\bibitem[{Guo et~al.(2024)Guo, Zhu, Yang, Xie, Dong, Zhang, Chen, Bi, Wu, Li et~al.}]{guo2024deepseek}
Daya Guo, Qihao Zhu, Dejian Yang, Zhenda Xie, Kai Dong, Wentao Zhang, Guanting Chen, Xiao Bi, Y~Wu, YK~Li, et~al. 2024.
\newblock Deepseek-coder: When the large language model meets programming--the rise of code intelligence.
\newblock \emph{arXiv preprint arXiv:2401.14196}.

\bibitem[{Josifoski et~al.(2023)Josifoski, Sakota, Peyrard, and West}]{josifoski2023exploiting}
Martin Josifoski, Marija Sakota, Maxime Peyrard, and Robert West. 2023.
\newblock Exploiting asymmetry for synthetic training data generation: Synthie and the case of information extraction.
\newblock \emph{arXiv preprint arXiv:2303.04132}.

\bibitem[{Kalyan et~al.(2018)Kalyan, Mohta, Polozov, Batra, Jain, and Gulwani}]{kalyan2018neural}
Ashwin Kalyan, Abhishek Mohta, Oleksandr Polozov, Dhruv Batra, Prateek Jain, and Sumit Gulwani. 2018.
\newblock Neural-guided deductive search for real-time program synthesis from examples.
\newblock \emph{arXiv preprint arXiv:1804.01186}.

\bibitem[{Lee et~al.(2024)Lee, Sim, Shin, Hwang, Seo, Park, Lee, Kim, and Kim}]{lee2024reasoning}
Seungpil Lee, Woochang Sim, Donghyeon Shin, Sanha Hwang, Wongyu Seo, Jiwon Park, Seokki Lee, Sejin Kim, and Sundong Kim. 2024.
\newblock \href {http://arxiv.org/abs/2403.11793} {Reasoning abilities of large language models: In-depth analysis on the abstraction and reasoning corpus}.

\bibitem[{Lee et~al.(2018)Lee, Heo, Alur, and Naik}]{lee2018accelerating}
Woosuk Lee, Kihong Heo, Rajeev Alur, and Mayur Naik. 2018.
\newblock Accelerating search-based program synthesis using learned probabilistic models.
\newblock \emph{ACM SIGPLAN Notices}, 53(4):436--449.

\bibitem[{Lei et~al.(2024)Lei, Lipovetzky, and Ehinger}]{lei2024generalized}
Chao Lei, Nir Lipovetzky, and Krista~A. Ehinger. 2024.
\newblock \href {http://arxiv.org/abs/2401.07426} {Generalized planning for the abstraction and reasoning corpus}.

\bibitem[{Li et~al.(2024{\natexlab{a}})Li, Zhou, Dong, Zhang, and Wang}]{arborist}
Xiang Li, Xiangyu Zhou, Rui Dong, Yihong Zhang, and Xinyu Wang. 2024{\natexlab{a}}.
\newblock \href {https://doi.org/10.1145/3632894} {Efficient bottom-up synthesis for programs with local variables}.
\newblock \emph{Proc. ACM Program. Lang.}, 8(POPL).

\bibitem[{Li et~al.(2024{\natexlab{b}})Li, Parsert, and Polgreen}]{li2024guiding}
Yixuan Li, Julian Parsert, and Elizabeth Polgreen. 2024{\natexlab{b}}.
\newblock \href {http://arxiv.org/abs/2403.03997} {Guiding enumerative program synthesis with large language models}.

\bibitem[{McCarthy(1960)}]{sexp}
John McCarthy. 1960.
\newblock \href {https://doi.org/10.1145/367177.367199} {Recursive functions of symbolic expressions and their computation by machine, part i}.
\newblock \emph{Commun. ACM}, 3(4):184–195.

\bibitem[{McCoy et~al.(2023)McCoy, Yao, Friedman, Hardy, and Griffiths}]{mccoy2023embers}
R~Thomas McCoy, Shunyu Yao, Dan Friedman, Matthew Hardy, and Thomas~L Griffiths. 2023.
\newblock Embers of autoregression: Understanding large language models through the problem they are trained to solve.
\newblock \emph{arXiv preprint arXiv:2309.13638}.

\bibitem[{Odena et~al.(2020)Odena, Shi, Bieber, Singh, Sutton, and Dai}]{odena2020bustle}
Augustus Odena, Kensen Shi, David Bieber, Rishabh Singh, Charles Sutton, and Hanjun Dai. 2020.
\newblock Bustle: bottom-up program synthesis through learning-guided exploration.
\newblock \emph{arXiv preprint arXiv:2007.14381}.

\bibitem[{OpenAI(2024)}]{openai_gpt4_2024}
OpenAI. 2024.
\newblock \href {https://openai.com/index/hello-gpt-4o/} {Hello gpt-4.0}.
\newblock Accessed: 2024-05-19.

\bibitem[{Osera and Zdancewic(2015)}]{osera2015type}
Peter-Michael Osera and Steve Zdancewic. 2015.
\newblock Type-and-example-directed program synthesis.
\newblock \emph{ACM SIGPLAN Notices}, 50(6):619--630.

\bibitem[{Reynolds et~al.(2019)Reynolds, Barbosa, N{\"o}tzli, Barrett, and Tinelli}]{reynolds2019cvc}
Andrew Reynolds, Haniel Barbosa, Andres N{\"o}tzli, Clark Barrett, and Cesare Tinelli. 2019.
\newblock cvc 4 sy: smart and fast term enumeration for syntax-guided synthesis.
\newblock In \emph{International Conference on Computer Aided Verification}, pages 74--83. Springer.

\bibitem[{Shi et~al.(2022{\natexlab{a}})Shi, Bieber, and Singh}]{tfcoder}
Kensen Shi, David Bieber, and Rishabh Singh. 2022{\natexlab{a}}.
\newblock \href {https://doi.org/10.1145/3517034} {Tf-coder: Program synthesis for tensor manipulations}.
\newblock \emph{ACM Trans. Program. Lang. Syst.}, 44(2).

\bibitem[{Shi et~al.(2022{\natexlab{b}})Shi, Dai, Ellis, and Sutton}]{shi2022crossbeam}
Kensen Shi, Hanjun Dai, Kevin Ellis, and Charles Sutton. 2022{\natexlab{b}}.
\newblock Crossbeam: Learning to search in bottom-up program synthesis.
\newblock \emph{arXiv preprint arXiv:2203.10452}.

\bibitem[{Tan and Motani(2023)}]{tan2023large}
John Chong~Min Tan and Mehul Motani. 2023.
\newblock Large language model (llm) as a system of multiple expert agents: An approach to solve the abstraction and reasoning corpus (arc) challenge.
\newblock \emph{arXiv preprint arXiv:2310.05146}.

\bibitem[{Udupa et~al.(2013)Udupa, Raghavan, Deshmukh, Mador-Haim, Martin, and Alur}]{udupa2013transit}
Abhishek Udupa, Arun Raghavan, Jyotirmoy~V Deshmukh, Sela Mador-Haim, Milo~MK Martin, and Rajeev Alur. 2013.
\newblock Transit: specifying protocols with concolic snippets.
\newblock \emph{ACM SIGPLAN Notices}, 48(6):287--296.

\bibitem[{Ugare et~al.(2024)Ugare, Suresh, Kang, Misailovic, and Singh}]{ugare2024improving}
Shubham Ugare, Tarun Suresh, Hangoo Kang, Sasa Misailovic, and Gagandeep Singh. 2024.
\newblock Improving llm code generation with grammar augmentation.
\newblock \emph{arXiv preprint arXiv:2403.01632}.

\bibitem[{Wang et~al.(2023)Wang, Zelikman, Poesia, Pu, Haber, and Goodman}]{wang2023hypothesis}
Ruocheng Wang, Eric Zelikman, Gabriel Poesia, Yewen Pu, Nick Haber, and Noah~D Goodman. 2023.
\newblock Hypothesis search: Inductive reasoning with language models.
\newblock \emph{arXiv preprint arXiv:2309.05660}.

\bibitem[{Wen et~al.(2024)Wen, Yin, Shi, Michalewski, Chaudhuri, and Polozov}]{wen2024grounding}
Yeming Wen, Pengcheng Yin, Kensen Shi, Henryk Michalewski, Swarat Chaudhuri, and Alex Polozov. 2024.
\newblock \href {http://arxiv.org/abs/2402.08073} {Grounding data science code generation with input-output specifications}.

\bibitem[{Wind(2020)}]{arc_kaggle_first}
Johan~Sokrates Wind. 2020.
\newblock \href {https://github.com/top-quarks/ARC-solution} {Arc kaggle competition, 1st place}.
\newblock Accessed: 2024-05-19.

\bibitem[{Xu et~al.(2023{\natexlab{a}})Xu, Khalil, and Sanner}]{xu2023graphs}
Yudong Xu, Elias~B Khalil, and Scott Sanner. 2023{\natexlab{a}}.
\newblock Graphs, constraints, and search for the abstraction and reasoning corpus.
\newblock In \emph{Proceedings of the AAAI Conference on Artificial Intelligence}, volume~37, pages 4115--4122.

\bibitem[{Xu et~al.(2023{\natexlab{b}})Xu, Li, Vaezipoor, Sanner, and Khalil}]{xu2023llms}
Yudong Xu, Wenhao Li, Pashootan Vaezipoor, Scott Sanner, and Elias~B Khalil. 2023{\natexlab{b}}.
\newblock Llms and the abstraction and reasoning corpus: Successes, failures, and the importance of object-based representations.
\newblock \emph{arXiv preprint arXiv:2305.18354}.

\bibitem[{Zhang et~al.(2023)Zhang, Dang, Peng, and Van~den Broeck}]{zhang2023tractable}
Honghua Zhang, Meihua Dang, Nanyun Peng, and Guy Van~den Broeck. 2023.
\newblock Tractable control for autoregressive language generation.
\newblock In \emph{International Conference on Machine Learning}, pages 40932--40945. PMLR.

\end{thebibliography}

\newpage
\appendix
\section{GPT4o Solutions for the Motivating Example}\label{app:gpt4o-solutions}

\begin{figure}[t]
\begin{lstlisting}
// Solution 1, occurs 6 times
if color_of(self) == GREY && is_neighbor(self, other) 
	then update_color(color_of(other))

// Solution 2, occurs 1 time
if is_neighbor(self, other) && color_of(other) == GREY 
	then update_color(color_of(other))

// Solution 3, occurs 1 time
if color_of(self) == GREY 
	then update_color(color_of(other))

// Solution 4, occurs 1 time
if not (color_of(self) == GREY) && is_neighbor(self, other) && color_of(other) == GREY
	then update_color(FUCHSIA)

// Solution 5, occurs 1 time
if size_of(self) == 4 then update_color(RED) ;
if size_of(self) == 4 && color_of(self) == GREY then update_color(FUCHSIA) ;
if size_of(self) == 4 && color_of(self) == BLUE then update_color(ORANGE) ;
if size_of(self) == 4 && color_of(self) == YELLOW then update_color(CYAN)
\end{lstlisting}
\caption{Ten samples from GPT4o for the motivating example in \autoref{fig:pbe-a}}\label{fig:gpt4o-solutions}		
\end{figure}

Recall the motivating example in \autoref{fig:pbe-a} where the task is to update the color of the grey objects to the color of their single-pixel neighbor.
As a reminder, the smallest correct solution to this task consists of the following rule:
\begin{lstlisting}
if color_of(self) == GREY && is_neighbor(self, x) && size_of(x) == MIN
	then update_color(color_of(x))
\end{lstlisting}
\autoref{fig:gpt4o-solutions} shows the programs we obtained by deduplicating 10 samples from GPT4o for this task.
The syntax of the solutions is slightly modified for readability;
our implementation uses a LISP-style s-expression syntax~\cite{sexp} to simplify parsing.

As you can see, the most frequent solution is almost correct,
except that it does not constrain the neighbor \T{other} to be of size 1;
this leads to the constraint being ambiguous (since every grey object has multiple neighbors of different colors),
in which case the program semantics is considered undefined.
That said, you can observe that the model consistently uses relevant components,
such as \T{color\_of}, \T{is\_neighbor}, and \T{update\_color},
which enables us to extract a useful PCFG from these solutions. 

When we increased the sample size to 125, GPT4o was able to produce one correct solution
(which is slightly larger than the minimal solution above):
\begin{lstlisting}
if color_of(self) == GREY && is_neighbor(self, other) && not (color_of(other) == GREY)
	then update_color(color_of(other))
\end{lstlisting}

\newpage
\section{LLM Prompt for the \arc Grammar}\label{app:llm-prompt}

\subsection{System Prompt}\label{app:llm-prompt-system}
The system prompt given to the LLM for \arc domain is shown in \autoref{fig:llm-prompt-system-arc}.
\begin{figure}
\begin{lstlisting}
You are an assistant chatbot with human-like perception, reasoning and learning capabilities.
You can solve tasks concisely, efficiently, and moreover, correctly.
Let's engage in perception and logic-based tasks.
You only output source code.
No explanations or any other text.
\end{lstlisting}
\caption{System prompt for \arc domain.}\label{fig:llm-prompt-system-arc}
\end{figure}

\subsection{User Prompt}\label{app:llm-prompt-user}
The full user prompt for the \arc domain is shown in \autoref{fig:llm-prompt-user-arc}.
It contains the domain-specific language, four in-context examples and the query for the test task.
    
\begin{figure}
\begin{lstlisting}[escapeinside={(`}{`)}, language=, breaklines=true, postbreak=\mbox{\textcolor{red}{$\hookrightarrow$}\space}]
You are an efficient assistant for logical reasoning and code generation.
You will help me solve a visual perception and reasoning task.
I will first provide you with the definition of a Domain Specific Language you will use for writing a solution for the task.
I will then present you with the description of the task that you will be tested in.
You will then respond to the queries I make regarding the solution of the task.

This is the definition of the DSL you will use to solve the task.
It is given as a context-free grammar in the EBNF format used by the Lark parser generator, with some informative comments about the semantics.
You will return a string that is parseable by the `program` non-terminal of the grammar.

```
library: "(" program* ")"

// Rules are executed one after another, in the order they appear.
// There could be no rules, in which case the program does nothing.
program: "(" "do" rule* ")"
...

<<< DSL IMPLEMENTATION IN LARK >>>

```
Now we continue with the visual perception and reasoning task.
The input for the task is a small number of pairs of grids of characters.
The value of each of the cells of the grids are the colors defined in the DSL, so we can think of grids as images.
Each pair of images correspond to an input-output example for an unknown program P.
For each pair, the program P is evaluated on the image grid and operates on the objects that appear in it.
The output of the program is then the output image.
The objects in the images are easy and natural to identify for humans, so there is no need to define them explicitly.
However you are able to abstract them correctly, and the DSL is interpreted with the same correct abstraction.

Now I will show you some demonstration tasks along with the output you would be expected to produce for each of them.

## DEMONSTRATION TASK 1

### INPUT
PAIR 1
INPUT GRID:
O O O O O O O O
O O O O O R O O
O R O O O R O R
O R R O O R O O
O O O O O O O O
O R R O O O O O
O R R O R R O O
O O O O O O O O

\end{lstlisting}
\caption{User prompt for \arc domain.}\label{fig:llm-prompt-user-arc}
\end{figure}

\begin{figure}
\begin{lstlisting}[escapeinside={(`}{`)}, language=, breaklines=true, postbreak=\mbox{\textcolor{red}{$\hookrightarrow$}\space}]
OUTPUT GRID:
O O O O O O O O
O O O O O Y O O
O Y O O O Y O Y
O Y Y O O Y O O
O O O O O O O O
O Y Y O O O O O
O Y Y O Y Y O O
O O O O O O O O

<<< ENCODING OF EXAMPLE PAIR 2 AND 3 OF DEMO TASK 1>>>

### EXPECTED OUTPUT
{
    "nl_description": "Recolor all objects to color Y",
    "code": <<< EXPECTED CODE IN DSL >>>
}

<<< MORE DEMONSTRATION TASKS (4 IN TOTAL) >>>

Now follows task you will be evaluated on.
Output the solution as a JSON object, which should contain both a natural language description of the solution and the solution written in the DSL.
The code should be parseable by the DSL grammar.
The JSON must have the following structure:

{
    "nl_description": "TO_BE_FILLED",
    "code": "TO_BE_FILLED"
}

## TEST TASK

PAIR 1
INPUT GRID:
O O R O O F O O O C
O O O O O O O O O O
O O O O X X X X O O
O O O O X X X X O O
O X X O X X X X O O
O X X O X X X X O O
O X X O O O O O O O
O X X O O O O X X X
O X X O O O O X X X
O O O O O O O X X X
OUTPUT GRID:
O O R O O F O O O C
O O O O O O O O O O
O O O O F F F F O O
O O O O F F F F O O
O R R O F F F F O O
O R R O F F F F O O
O R R O O O O O O O
O R R O O O O C C C
O R R O O O O C C C
O O O O O O O C C C

<<< REST OF THE I/O EXAMPLES OF TEST TASK >>>
\end{lstlisting}
\end{figure}

\newpage
\section{Experimental results with LLMs \deepseek and \gptthree}\label{app:other-models}
\begin{figure}[h]
\begin{subfigure}{.5\textwidth}
\centering
\includegraphics[width=\textwidth]{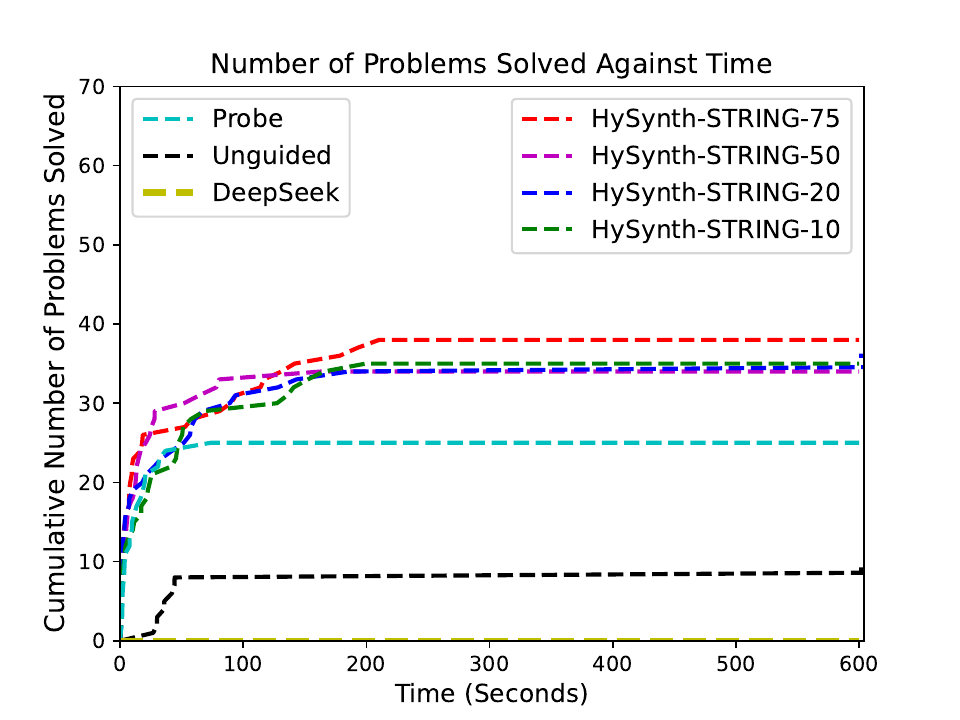}
\caption{\tool-\stringbench results with \deepseek} \label{fig:string-deepseek}
\end{subfigure}%
\begin{subfigure}{.5\textwidth}
\centering
\includegraphics[width=\textwidth]{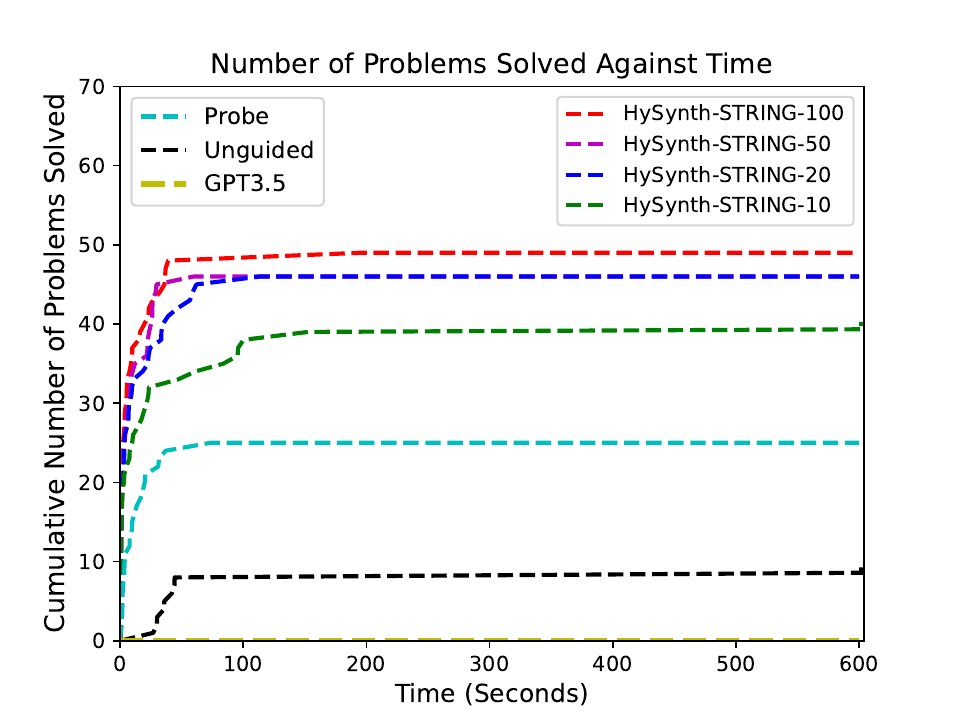}
\caption{\tool-\stringbench results with \gptthree}
\label{fig:string-gpt3}
\end{subfigure}%

\begin{subfigure}{.5\textwidth}
\centering
\includegraphics[width=\textwidth]{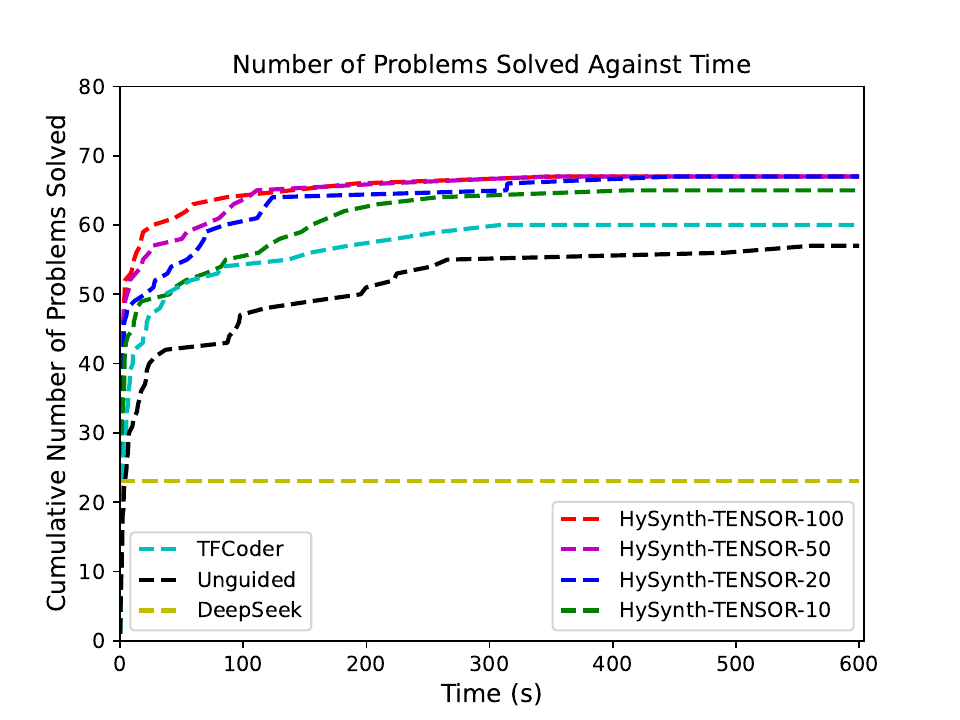}
\caption{\tool-\tensorbench results with \deepseek} \label{fig:tensor-deepseek}
\end{subfigure}%
\begin{subfigure}{.5\textwidth}
\centering
\includegraphics[width=\textwidth]{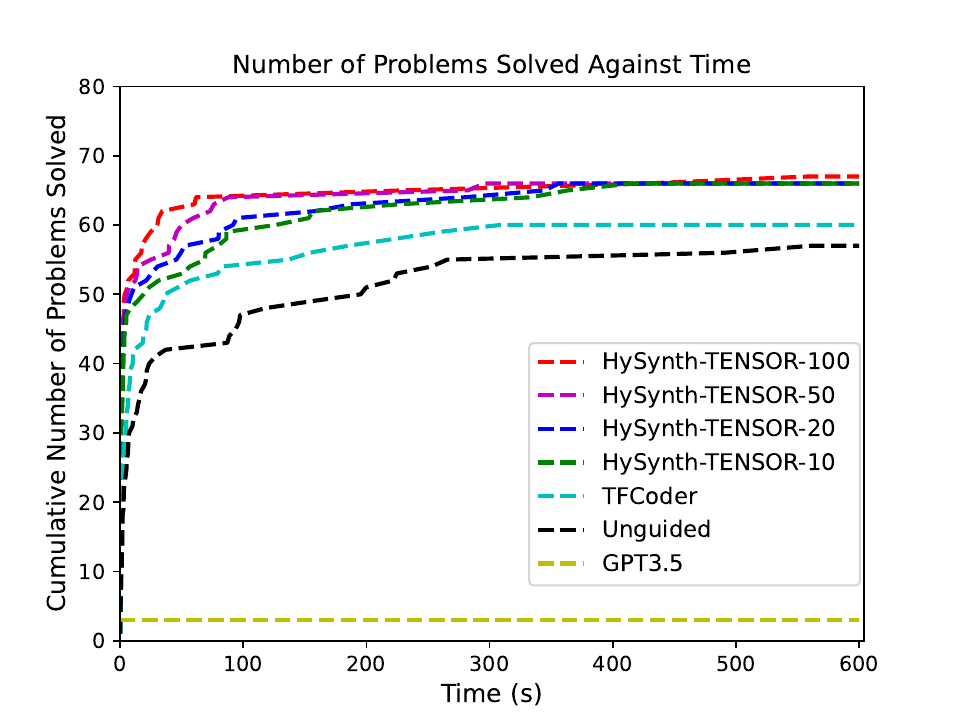}
\caption{\tool-\tensorbench results with \gptthree}\label{fig:tensor-gpt3}
\end{subfigure}
\caption{\tool-\stringbench and \tool-\tensorbench evaluation results with \deepseek and \gptthree.}
\label{fig:other-models}
\end{figure}

\newpage

\section{LLM Prompt for the \tensorbench Grammar}
The system and user prompt for \tensorbench domain are in \autoref{fig:sysprompt-tensor} and \autoref{fig:userprompt-tensor}.
\begin{figure}
\begin{lstlisting}
You are a coding assistant. Be precise and terse.
You will be provided a list of tensorflow operators, a task description, and some input/output examples.
Your task is to generate the body of a python function that will transform the input to the output.
Only use the operators provided in the list.
Your answer should be as short as possible while still being correct.
Make sure to only generate python code.
\end{lstlisting}
\caption{System prompt for \tensorbench domain.}\label{fig:sysprompt-tensor}
\end{figure}

\begin{figure}
\begin{lstlisting}
[TENSORFLOW OPERATORS]
<<< see appendix E >>>

[TASK DESCRIPTION]
index into the tensor

[INPUTS]
[[ 5.  2.]
 [ 1.  3.]
 [ 0. -1.]]


[OUTPUTS]
[[[ 5.  5.]
  [ 1.  1.]
  [ 0.  0.]]

 [[ 2.  2.]
  [ 3.  3.]
  [-1. -1.]]]

[PROGRAM]
def transform(in1):
\end{lstlisting}
\caption{User prompt for \tensorbench domain}\label{fig:userprompt-tensor}
\end{figure}

\newpage
\section{LLM Prompt for \stringbench}

The system and user prompt for \stringbench domain are in \autoref{fig:sysprompt-string} and \autoref{fig:userprompt-string}.

\begin{figure}
\begin{lstlisting}[escapeinside={(`}{`)}, language=, breaklines=true, postbreak=\mbox{\textcolor{red}{$\hookrightarrow$}\space}]
You are a coding assistant. Be precise and terse.
You will be given a SyGuS grammar, a natural language specification, and a set of input-output examples.
Your task is to complete the provided function definition with an implementation that is correct according to the grammar, specification, and examples.
Your answer should be as short as possible while still being correct.
Make sure that your answer is a valid s-expression.
\end{lstlisting}
\caption{System prompt for \stringbench domain}\label{fig:sysprompt-string}
\end{figure}

\begin{figure}
\begin{lstlisting}[escapeinside={(`}{`)}, language=, breaklines=true, postbreak=\mbox{\textcolor{red}{$\hookrightarrow$}\space}]
[GRAMMAR]
(synth-fun f ((_arg_0 String)) String ((Start String (ntString)) (ntString String (_arg_0 "" " " "BRD" "DRS" "LDS" "Branding" "Direct Response" "Leads" "=" "/" "in" "_" "9" "." "microsoft" "windows" "apple" "mac" "-" "1" "2" "3" "4" "5" "6" "7" "8" "0" "," "<" ">" "/n" "%" "b" "apple" "bananas" "strawberries" "oranges" "LLC" "Inc" "Corporation" "Enterprises" "Company" "(" ")" "+" "name" "," (str.++ ntString ntString) (str.replace ntString ntString ntString) (str.at ntString ntInt) (int.to.str ntInt) (ite ntBool ntString ntString) (str.substr ntString ntInt ntInt))) (ntInt Int (-1 1 2 3 4 5 6 7 8 9 0 1 0 -1 (+ ntInt ntInt) (- ntInt ntInt) (str.len ntString) (str.to.int ntString) (ite ntBool ntInt ntInt) (str.indexof ntString ntString ntInt))) (ntBool Bool (true false (= ntInt ntInt) (str.prefixof ntString ntString) (str.suffixof ntString ntString) (str.contains ntString ntString)))))

[NATURAL LANGUAGE SPECIFICATION]
; https=//exceljet.net/formula/get-top-level-domain-tld

[EXAMPLES]
www.domain.com -> com
mail.net -> net
www.amazon.co.uk -> uk

[SOLUTION]
(define-fun f (_arg_0 String) String
\end{lstlisting}
\caption{User message for \stringbench}\label{fig:userprompt-string}
\end{figure}

\section{The Full \stringbench Grammar}\label{sec:string-grammar}
\begin{figure}[]
	\begin{center}
		\begin{align*}
		\nonterm{Start} &\rightarrow && S &&\\
		S &\rightarrow &&\ \scode{arg0}\ |\  \scode{arg1} |\ \dots && \quad\text{string variables}\\
		&\mid&& \scode{lit-1 } |\ \scode{lit-2 } |\ \dots && \quad\text{string literals}\\
		&\mid &&\ (\scode{replace}\ S\ S\ S) && \quad\text{\scode{replace s x y} replaces first occurrence of \scode{x} in \scode{s} with \scode{y}}\\
		&\mid &&\ (\scode{concat}\ S\ S) && \quad\text{\scode{concat x y} concatenates \scode{x} and \scode{y}}\\
		&\mid &&\ (\scode{substr}\ S\ I\ I) && \quad\text{\scode{substr x y z} extracts substring of length \scode{z}, from index \scode{y}}\\
		&\mid &&\ (\scode{ite}\ B\ S\ S) && \quad\text{\scode{ite x y z} returns \scode{y} if \scode{x} is true, otherwise \scode{z}}\\
		&\mid &&\ (\scode{int.to.str}\ I) && \quad\text{\scode{int.to.str x} converts int \scode{x} to a string}\\	
		&\mid &&\ (\scode{at}\ S\ I) && \quad\text{\scode{at x y} returns the character at index \scode{y} in string {x}}\\
		B &\rightarrow &&\ \scode{true} \mid \scode{false} && \quad\text{bool literals}\\
		&\mid &&\ (\scode{=}\ I\ I) && \quad\text{\scode{= x y} returns true if \scode{x} equals \scode{y}}\\	
		&\mid &&\ (\scode{contains}\ S\ S) && \quad\text{\scode{contains x y} returns true if \scode{x} contains \scode{y}}\\	
		&\mid &&\ (\scode{suffixof}\ S\ S) && \quad\text{\scode{suffixof x y} returns true if \scode{x} is the suffix of \scode{y}}\\
		&\mid &&\ (\scode{prefixof}\ S\ S) && \quad\text{\scode{prefixof x y} returns true if \scode{x} is the prefix of \scode{y}}\\
		I &\rightarrow &&\ \scode{arg0}\ |\  \scode{arg1} |\ \dots && \quad\text{int variables}\\
		&\mid&& \scode{lit-1 } |\ \scode{lit-2 } |\ \dots && \quad\text{int literals}\\
		&\mid &&\ (\scode{str.to.int}\ S) && \quad\text{\scode{str.to.int x} converts string x to a int}\\	
		&\mid &&\ (\scode{+}\ I\ I) && \quad\text{\scode{+ x y} sums \scode{x} and \scode{y}}\\
		&\mid &&\ (\scode{-}\ I\ I) && \quad\text{\scode{- x y} subtracts \scode{y} from \scode{x}}\\
		&\mid &&\ (\scode{length}\ S) && \quad\text{\scode{length x} returns length of \scode{x}}\\
		&\mid &&\ (\scode{ite}\ B\ I\ I) && \quad\text{\scode{ite x y z} returns \scode{y} if \scode{x} is true, otherwise \scode{z}}\\
		&\mid &&\ (\scode{indexof}\ S\ S\ I) && \quad\text{\scode{indexof x y z} returns index of \scode{y} in \scode{x}, starting at index \scode{z}}\\
		\end{align*}
	\end{center}
	\caption{The full \sygus \stringbench grammar of the \probe benchmark suite. Integer and string variables and constants change per benchmark. Some benchmark files contain a reduced grammar.}\label{fig:eval-grammar}
\end{figure}%
The full grammar for the \stringbench domain is detailed in \autoref{fig:eval-grammar}.

\newpage
\section{The Full \tensorbench Grammar}\label{sec:tensor-grammar}

\begin{minipage}{\linewidth} 
\begin{figure}[H]
\begin{lstlisting}
General TensorFlow functions:
-----------------------------
tf.abs(x)
tf.add(x, y)
tf.add_n(inputs)
tf.argmax(input, axis)
tf.argmin(input, axis)
tf.argsort(values, axis, stable=True)
tf.argsort(values, axis, direction='DESCENDING', stable=True)
tf.boolean_mask(tensor, mask)
tf.broadcast_to(input, shape)
tf.cast(x, dtype)
tf.clip_by_value(t, clip_value_min, clip_value_max)
tf.concat(values, axis)
tf.constant(value)
tf.constant(value, dtype)
tf.divide(x, y)
tf.equal(x, y)
tf.exp(x)
tf.expand_dims(input, axis)
tf.eye(num_rows)
tf.eye(num_rows, num_columns)
tf.eye(num_rows, dtype)
tf.fill(dims, value)
tf.gather(params, indices)
tf.gather(params, indices, axis, batch_dims)
tf.gather_nd(params, indices)
tf.gather_nd(params, indices, batch_dims)
tf.greater(x, y)
tf.greater_equal(x, y)
tf.math.bincount(arr)
tf.math.ceil(x)
tf.math.count_nonzero(input)
tf.math.count_nonzero(input, axis)
tf.math.cumsum(x, axis)
tf.math.cumsum(x, axis, exclusive=True)
tf.math.divide_no_nan(x, y)
tf.math.floor(x)
tf.math.log(x)
tf.math.negative(x)
tf.math.reciprocal(x)
tf.math.reciprocal_no_nan(x)
tf.math.segment_max(data, segment_ids)
tf.math.segment_mean(data, segment_ids)
tf.math.segment_min(data, segment_ids)
tf.math.segment_prod(data, segment_ids)
tf.math.segment_sum(data, segment_ids)
tf.math.squared_difference(x, y)
tf.math.top_k(input, k)
tf.math.unsorted_segment_max(data, segment_ids, num_segments)
tf.math.unsorted_segment_mean(data, segment_ids, num_segments)
tf.math.unsorted_segment_min(data, segment_ids, num_segments)
tf.math.unsorted_segment_prod(data, segment_ids, num_segments)
tf.math.unsorted_segment_sum(data, segment_ids, num_segments)
\end{lstlisting}
\caption{List of \tensorflow operations as used in \tfcoder.}
\end{figure}
\end{minipage}

\begin{minipage}{\linewidth} 
\begin{figure}[H]
\begin{lstlisting}
tf.matmul(a, b)
tf.maximum(x, y)
tf.minimum(x, y)
tf.multiply(x, y)
tf.not_equal(x, y)
tf.one_hot(indices, depth)
tf.ones(shape)
tf.ones_like(input)
tf.pad(tensor, paddings, mode='CONSTANT')
tf.pad(tensor, paddings, mode='CONSTANT', constant_values)
tf.pad(tensor, paddings, mode='REFLECT')
tf.pad(tensor, paddings, mode='SYMMETRIC')
tf.range(start)
tf.range(start, limit, delta)
tf.reduce_any(input_tensor, axis)
tf.reduce_max(input_tensor)
tf.reduce_max(input_tensor, axis)
tf.reduce_mean(input_tensor)
tf.reduce_mean(input_tensor, axis)
tf.reduce_min(input_tensor)
tf.reduce_min(input_tensor, axis)
tf.reduce_prod(input_tensor, axis)
tf.reduce_sum(input_tensor)
tf.reduce_sum(input_tensor, axis)
tf.reshape(tensor, shape)
tf.reverse(tensor, axis)
tf.roll(input, shift, axis)
tf.round(x)
tf.searchsorted(sorted_sequence, values, side='left')
tf.searchsorted(sorted_sequence, values, side='right')
tf.sequence_mask(lengths)
tf.sequence_mask(lengths, maxlen)
tf.shape(input)
tf.sign(x)
tf.sort(values, axis)
tf.sort(values, axis, direction='DESCENDING')
tf.sqrt(x)
tf.square(x)
tf.squeeze(input)
tf.squeeze(input, axis)
tf.stack(values, axis)
tf.subtract(x, y)
tf.tensordot(a, b, axes)
tf.tile(input, multiples)
tf.transpose(a)
tf.transpose(a, perm)
tf.unique_with_counts(x)
tf.unstack(value, axis)
tf.where(condition)
tf.where(condition, x, y)
tf.zeros(shape)
tf.zeros_like(input)

SparseTensor functions:
-----------------------
tf.SparseTensor(indices, values, dense_shape)
tf.sparse.add(a, b)
tf.sparse.concat(axis, sp_inputs)
tf.sparse.expand_dims(sp_input, axis)
\end{lstlisting}
\end{figure}
\end{minipage}

\begin{minipage}{\linewidth} 
\begin{figure}[H]
\begin{lstlisting}
tf.sparse.from_dense(tensor)
tf.sparse.maximum(sp_a, sp_b)
tf.sparse.minimum(sp_a, sp_b)
tf.sparse.reduce_max(sp_input, axis, output_is_sparse)
tf.sparse.reduce_sum(sp_input, axis, output_is_sparse)
tf.sparse.reset_shape(sp_input)
tf.sparse.reshape(sp_input, shape)
tf.sparse.retain(sp_input, to_retain)
tf.sparse.slice(sp_input, start, size)
tf.sparse.split(sp_input, num_split, axis)
tf.sparse.to_dense(sp_input)
tf.sparse.to_dense(sp_input, default_value)
tf.sparse.to_indicator(sp_input, vocab_size)
tf.sparse.transpose(sp_input)
tf.sparse.transpose(sp_input, perm)

Python-syntax operations:
-------------------------
IndexingAxis1Operation: arg1[:, arg2]
IndexingOperation: arg1[arg2]
PairCreationOperation: (arg1, arg2)
SingletonTupleCreationOperation: (arg1,)
SlicingAxis0BothOperation: arg1[arg2:arg3]
SlicingAxis0LeftOperation: arg1[arg2:]
SlicingAxis0RightOperation: arg1[:arg2]
SlicingAxis1BothOperation: arg1[:, arg2:arg3]
SlicingAxis1LeftOperation: arg1[:, arg2:]
SlicingAxis1RightOperation: arg1[:, :arg2]
TripleCreationOperation: (arg1, arg2, arg3)
\end{lstlisting}
\end{figure}
\end{minipage}

\newpage

\section{The Full \arc DSL}\label{app:arc-grammar}
The full grammar of our \arc DSL is shown in \autoref{fig:arcgrammar}.

\begin{figure}[t]
  \begin{center}
  \begin{align*}
  \nonterm{Rule}  &\rightarrow && \T{if}\ \nonterm{Filter}\ \T{then}\ \nonterm{Transforms}\\
\nonterm{Transforms} &\rightarrow && \nonterm{Transform} \mid \nonterm{Transform} \ \T{;} \ \nonterm{Transforms} \\
  \nonterm{Filter} &\rightarrow && \nonterm{Atom} \mid \T{not}\ \nonterm{Atom}\mid \nonterm{Atom}\ \T{&&}\ \nonterm{Filter} \mid  \nonterm{Atom}\ \T{||}\ \nonterm{Filter}\\
  \nonterm{Atom} &\rightarrow && \nonterm{Color}\ \T{==}_c\ \nonterm{Color} \mid \nonterm{Size}\ \T{==}_s\ \nonterm{Size} \mid \nonterm{Degree}\ \T{==}_d\ \nonterm{Degree}
  \mid \nonterm{Width}\ \T{==}_w\ \nonterm{Width}
  \mid \nonterm{Height}\ \T{==}_h\ \nonterm{Height} &&\\
  &\mid && \nonterm{Shape}\ \T{==}_{S} \nonterm{Shape}
  \mid \nonterm{Row}\ \T{==}_r\ \nonterm{Row}\
  \mid \nonterm{Column}\ \T{==}_C\ \nonterm{Column}
  \mid \T{is\_neighbor}\ \T{(}\nonterm{Obj, Obj}\T{)}\\
  \nonterm{Transform} &\rightarrow && \T{update\_color(}\nonterm{Color} \T{)} \mid \T{move(}\nonterm{Dir} \T{)} \mid \T{move\_max(}\nonterm{Dir}\T{)} \mid
  \T{extend(}\nonterm{Dir, Overlap}\T{)}&&\\
  &\mid && \T{rotate(}\nonterm{Angle} \T{)} \mid \T{fill\_rectangle(}\nonterm{Color, Overlap} \T{)} \mid \T{hollow\_rectangle(}\nonterm{Color}\T{)}\\
  &\mid && \T{mirror(}\nonterm{Axis}\T{)}  \mid \T{add\_border(}\nonterm{Color} \T{)} \mid \T{flip(}\nonterm{Axis}\T{)} \mid \T{NoOp} \\
  \nonterm{Obj} &\rightarrow && \T{self} \mid \T{x} \mid \T{y} \mid \dots\\
  \nonterm{Color} &\rightarrow && \T{color\_of(}\nonterm{Obj} \T{)} \mid \T{GREY} \mid \T{RED} \mid \T{BLACK} \mid \T{BLUE} \mid \T{YELLOW} \mid \T{ORANGE} \mid \T{BROWN} 
  \mid \T{GREEN} \mid \T{GREY} \mid \T{FUCHSIA} \dots\\
  \nonterm{Dir} &\rightarrow && \T{dir\_of(}\nonterm{Obj} \T{)} \mid \T{UP} \mid \T{DOWN} \mid \T{LEFT} \mid \T{RIGHT}
  \mid \T{UPLEFT} \mid \T{DOWNLEFT} \mid \T{UPRIGHT} \mid \T{DOWNRIGHT} \dots\\
\nonterm{Axis} &\rightarrow && \T{axis\_of(}\nonterm{Obj} \T{)} \mid \T{VERTICAL}\ \mid \T{HORIZONTAL} \mid \T{LEFTDIAGONAL} \mid \T{RIGHTDIAGONAL} \dots\\
\nonterm{Overlap} &\rightarrow && \T{TRUE} \mid \T{FALSE}\\
\nonterm{Angle} &\rightarrow && \T{90}\ \mid \T{180} \mid \T{270}\ &&\\
\nonterm{Size} &\rightarrow && \T{size\_of(}\nonterm{Obj} \T{)} \mid \T{MIN} \mid \T{MAX} \mid \dots\\
\nonterm{Degree} &\rightarrow && \T{degree\_of(}\nonterm{Obj} \T{)} \mid \T{MIN} \mid \T{MAX} \mid \dots\\
\nonterm{Width} &\rightarrow && \T{width\_of(}\nonterm{Obj} \T{)} \mid \T{MIN} \mid \T{MAX} \mid \dots\\
\nonterm{Height} &\rightarrow && \T{height\_of(}\nonterm{Obj} \T{)} \mid \T{MIN} \mid \T{MAX} \mid \dots\\
\nonterm{Column} &\rightarrow && \T{column\_of(}\nonterm{Obj} \T{)} \mid \T{MIN} \mid \T{MAX} \mid \dots\\
\nonterm{Row} &\rightarrow && \T{row\_of(}\nonterm{Obj} \T{)} \mid \T{MIN} \mid \T{MAX} \mid \dots\\
\nonterm{Shape} &\rightarrow && \T{shape\_of(}\nonterm{Obj} \T{)} \mid \T{ENCLOSED} \mid \T{SQUARE} \mid \dots\\
  \end{align*}
  \end{center}
  \caption{The full grammar for our \arc DSL, object specific parameters like size, degree change per benchmark.}\label{fig:arcgrammar}
\end{figure}


\section{Detailed Prompt Settings}
For \arc, we sample completions with temperature 1 and 4000 max tokens. For \tensorbench, we use temperature 1 and 4000 max tokens. For \sygus, we use temperature 0.5 and 4000 max tokens. We use the same settings for all three LLMs. When prompting \gptfouro, we set \T{response_type} to JSON.

\section{Broader Research Impacts}
Our technique presents a powerful strategy for harnessing both syntactically valid and invalid outputs from an LLM to learn a surrogate model.
Incorporating hallucinatory outputs -- often erroneous generated by the model, allows us to extract insights that are discarded in standard practices.
Our approach mitigates the need for large-scale sampling of completions from LLMs, promoting a more efficient and effective utilization of these models, saving resources.
In addition to improving the cost effectiveness of using LLMs, it also opens up new avenues for enhancing model robustness and adaptability across different domains.

\newpage
\section{The \arc Synthesis Algorithm}\label{app:arc-algo}

\paragraph{Overall Synthesis Algorithm}
The overall synthesis algorithm takes as input a set of input-output grids $\examples$, along with grammars $\grammar_t$ and $\grammar_f$.
We sample candidate solutions from an LLM by constructing a prompt using $\examples$.
These solutions are used to initialize the weights of production rules in the transform and filter grammars, $\grammar_t$ and $\grammar_f$, respectively.
We optimize the search by using a divide and conquer approach: first, a \textproc{Transform-Search} procedure searches for transforms, mapping each to its correctly transformed objects in $\oset$.
Following this, a search for filters is initiated using the \textproc{Filter-Search} procedure.
%
If a filter is found for each transform, the algorithm terminates and returns $\mset$, which maps each transform to its corresponding filter.
The algorithm described above terminates after the first solution is found, but we keep searching for a smaller set of transforms~\cite{alur2017scaling}.
\begin{algorithm}[t]
\small
\caption{\arc Synthesis Algorithm}\label{arc:algo}
\begin{algorithmic}[1]
\Require{A set of input-output example grids $\examples$, transform grammar $\grammar_t$ and filter grammar $\grammar_f$}
\Ensure{A solution map  $\mset$ from each transform to the corresponding filter}
\Procedure{\tool-\arc}{$\examples$, $\grammar_p$, $\grammar_t$}
\State $\level, \bank \gets 0, \emptyset$ \Comment{Initialize search state}

\State $ \llmSols \gets \textproc{Llm}(\examples)$
\Comment{Sample solutions from the LLM}
\State $\grammar_p, \grammar_t \gets \textproc{Init}(\grammar_p,  \llmSols), \textproc{Init}(\grammar_t,  \llmSols)$  \Comment{Initialize both PCFGs using LLM solutions}

\While{not timeout}                  
\State $\oset \gets \textproc{Transform-Search}(\grammar_t, \examples)$																				\Comment{Synthesize transforms that cover all objects}
\State $\mset \gets \textproc{Filter-Search}(\grammar_f, \examples, \oset)$ 
\Comment{Synthesize filters for the above transforms}
\If{$\forall (t, f) \in \mset, f \neq \bot$} 	\Comment{Found a filter for each transform}
\State \Return $\mset$							\Comment{Return the complete solution}
\EndIf
\EndWhile
\EndProcedure
\end{algorithmic}
\end{algorithm}

\paragraph{Transform Search Algorithm}

\begin{algorithm}[t!]
	\small
	\caption{Transform Synthesis Algorithm}\label{talgo}
	\begin{algorithmic}[1]
		\Require{PCFG $\grammar_t$ and input-output grids  $\examples$}
		\Ensure{A map $\oset$ from transforms to correctly changed objects}
		\Procedure{Transforms-Search}{$\grammar_t, \examples$}
		\State $\level, \bank, \exec \gets 0,\emptyset,\emptyset$
		\Comment{Initialize search state}
		\While{$\level \leq \llim$}
		\For{$\transform \in \textproc{New-Transforms}(\grammar_t, \level, \bank)$}  \Comment{For all transforms with cost $\level$}
\State $\eval \gets \left\{{\sem{\transform}(\omega_i)} \ \middle| \ \langle i, o \rangle \in \examples, \omega_i \in i \right\}$
\Comment{Apply transform on objects in input grids from $\examples$}
\If{$\eval \cap \bigcup_{\langle i, o \rangle \in \examples} \{\omega_o \mid \omega_o \in o\} \neq \emptyset$}
\Comment{$\transform$ covers a subset of objects}
\State $\oset[\transform] \gets \eval$ \Comment{Store the transform and objects covered by it}
\ElsIf{$\eval \in \exec$}
\State{\textbf{continue}}  \Comment{$\transform$ is observationally equivalent to another transform in $\bank$}
\EndIf
\If{$\bigcup_{\transform \in \oset} \oset[\transform] = \bigcup_{\langle i, o \rangle \in \examples} \{ \omega_o \mid \omega_o \in o \}$}
\Comment{All objects are correctly transformed}
		\State \textbf{return} $\oset$
		\EndIf
		\State $\bank[\level] \gets \bank[\level] \cup \{\transform\}$ \Comment{Add transform to the bank, indexed by cost for later search}
		\State $\exec \gets \exec \cup \eval$                             \Comment{Cache evaluation result}
		\EndFor
		\State $\level \gets \level + 1$
		\EndWhile
		\State \textbf{return} $\bot$    \Comment{Cost limit reached}
		\EndProcedure
\end{algorithmic}
\end{algorithm}
The transform synthesis algorithm in Algorithm~\ref{talgo} takes as input a PCFG $\grammar_t$ and $\examples$.
It enumerates transforms in the order of increasing discrete costs according to $\grammar_t$.

The algorithm starts with the following initial state:
1) a cost level ($\level$) equal to 0 in order to keep track of the current cost during enumeration,
2) a program bank ($\bank$) that indexes the enumerated transforms by their cost for efficient retrieval, and
3) an evaluation cache ($\exec$) that stores the result of all evaluated transforms within $\bank$.
At each iteration, the algorithm explores the space of all new transforms generated by the $\textproc{New-Transforms}$ procedure for the current cost level.

On line 5 in Algorithm~\ref{talgo}, the enumerated transform $\transform$ is applied to each object in the input grids from $\examples$.
If $\transform$ correctly transforms a subset of objects, $\transform$ and the objects covered by it are stored in map $\oset$ indexed by the transform (line 7).
%
%
When the transforms in $\oset$ cover all grid objects, $\oset$ is returned (line 10-11).
For transforms with objects bound by a filter, such as in \scode{update\_color(color\_of(other))}, we consider all possible values (of color) that could be assigned and yield concrete transforms corresponding to each of those assignments.
\paragraph{Filter Search Algorithm}
The filter search algorithm takes as input a filter PCFG $\grammar_f$, $\examples$, and
the map $\oset$ returned by the transform search in Algorithm~\ref{talgo}.
The filter search proceeds in a similar manner as the transforms search wherein it enumerates filters in the order of increasing cost as per the PCFG  $\grammar_f$.
It initiates a new search to find a filter for each transform in $\oset$.
Each enumerated filter expression is evaluated on all objects in the input grids.
If the objects for which the filter is True are the same as the objects covered by the transform, we have found a filter for this transform.
Once a filter is found for each of the transforms in $\oset$, we return the solution map $\mset$.

\end{document}